\documentstyle[sprocl]{article}

\input{psfig}
\def\Journal#1#2#3#4{{#1} {\bf #2}, #3 (#4)}

\def\PRL{\em Phys. Rev. Lett.}

\def\be{\begin{equation}}
\def\ee{\end{equation}}
\def\bea{\begin{eqnarray}}
\def\eea{\end{eqnarray}}
\begin{document}

\markright{CLNS 96/1453}
\title{EXPERIMENTAL ASPECTS OF THE STANDARD MODEL:  A SHORT COURSE FOR
THEORISTS}

\author{ PERSIS S. DRELL }

\address{Cornell University, Newman Laboratory  \\ Ithaca,
NY 14853-5001}


\maketitle\abstracts{Abstract}
This is a series of lectures intended to introduce high energy theorists to
the marvels of the Standard Model from an experimentalist's point of view.
 
\section{Introduction}

The subject of the 1996 TASI summer school was ``Fields, Strings, and Duality" so
it is surprising, perhaps, to see a series of lectures by
an experimentalist included
as part of the school.  I wrote these lectures because
I believe with deep conviction that physics is an experimental science.  
It is the goal of physics to describe how the world works at its most basic
and fundamental level.  The students at TASI 1996 were all, for the most
part, embarking on careers in theoretical physics, and I felt that I could
not pass up the opportunity to teach them a little about how experimentalists
view the world.  The test of all scientific knowledge is experiment and the
most beautiful and elegant theoretical model has no lasting value if it cannot
be used to describe the results of experiments.

It is quite possible that the theories discussed at TASI 96 will not be
tested experimentally for decades.  On the other hand, perhaps they will 
be able to address anomalies in current data sets or in data that will
be collected in the first half of the twenty-first century.  Without an
accurate crystal ball, I personally believe that it is extraordinarily important
that theorists and experimentalists at least be able to talk to each other. 
We need to have some common language, and some understanding of 
each other's techniques 
and the scope of the problems we are each trying to address.

As theories become more formal and mathematical and experiments 
become more complex and
difficult, theory and experiment grow apart.  It will take effort on the
part of both theorists and experimentalists  to stay in touch with each other.
With these lectures, I hope to provide theory students with some tools to make
that task easier and to motivate them to put in the effort
required to forge the communication links with their experimental colleagues.

The subjects covered in these lectures are organized as follows:
\begin{itemize}
\item In Section 2, 
I will briefly discuss the Standard Model.  I will list the 
free parameters of the Standard Model in the language of an experimentalist,
and I will then describe two of the experiments that were done in the 1970's
that convinced us that the Standard Model provides an accurate  description of
the world.
\item In Section 3, I will discuss the anatomy of an experimental result.  This
section is meant to introduce a student of theory to the main tools of the
experimentalist; how a measurement is made, and what kinds of questions should
be asked when trying to decide whether to believe a result or not. Contrary
to what you may have heard, experimentalists are occasionally wrong. I'll end
this section with a discussion of the pitfalls that occasionally snare experimentalists
and theorists alike.
\item In Section 4, I will discuss experiments that define the Standard Model
by measuring some of the free parameters of the theory. I will discuss how to
measure a coupling constant, a gauge boson mass, a Yukawa coupling, and a quark
mixing angle.
\item The final section will concentrate on experiments that are testing the
validity of the Standard Model. I will talk about searches for the Higgs and
precision measurements made at the $Z$ pole. I will try to summarize the
status of our current understanding of the Standard Model and what are the rogue results
awaiting confirmation that could be our first hints of new physics beyond the
Standard Model. I will end the lectures with a discussion of experiments
of the future. Where will the data be coming from over the next few decades
and what physics will they hope to address?
\end{itemize}

\section{Standard Model Basics}

\subsection{Overview of the Standard Model}

These lectures assume familiarity with the Standard Model of electroweak
interactions (hereafter referred to as SM). The SM is a gauge theory where the
requirement of local gauge invariance under chiral isospin transformations 
results in the minimal
couplings to the matter fields. The gauge bosons of the theory acquire a mass
via the Higgs mechanism which leads to the prediction of a massive scalar boson
in the model which is yet to be discovered
experimentally. The fermions in the model acquire mass via a
Yukawa coupling to this Higgs field. It is worth keeping in mind that the
process by which the gauge bosons acquire a mass derives from the very
elegant procedure of spontaneous symmetry breaking and the existence of a finite
vacuum expectation value for the Higgs field, so we at least think
we understand the origins of the gauge boson masses. The fermion masses
are introduced in a totally ad hoc fashion into the model.

The correct gauge group to describe nature is not predicted by the model. The 
simplest choice consistent with existing phenomenology was suggested in 1968
by Weinberg to be $SU(2)_L \times U(1)$ and reflected the known $V-A$ nature of
the charged weak interactions. This choice is consistent with all experimental
data to date but keep in mind that with this choice, the most striking feature
of the weak interactions is simply inserted by hand. 
Once the gauge group is known, there are many free parameters in the model that
must be determined. 
These are listed in Table~\ref{tab:SM}.

\begin{table}[t]
\caption{Free parameters in the Standard Model of electroweak interactions.
\label{tab:SM}}

\vspace{0.4cm}
\begin{center}
\begin{tabular}{|c|c|c|}
\hline
 & & \\
 & Theorists & Experimentalists \\
 &  & \\  \hline
 &  & \\ 
Gauge Couplings and & $g, g', g_3$  & $\alpha_{EM}, G_F, \alpha_3$  \\
Parameters of Higgs Field & $v, \mu$  & $M_Z,M_H$  \\
 &  & \\ \hline
 & \multicolumn{2}{c|} {}  \\ 
Fermion& \multicolumn{2}{c|} {$m_e,m_\mu,m_\tau,m_{\nu_e},
m_{\nu_\mu},m_{\nu_\tau}$ } \\
Masses & \multicolumn{2}{c|} {$m_u,m_c,m_t,m_d,m_s,m_b$ }\\        
 &\multicolumn{2}{c|}  {} \\ \hline
 &\multicolumn{2}{c|} {}  \\ 
Quark & \multicolumn{2}{c|} { $V_{ud}, V_{us}, V_{ub}$} \\
Mixing & \multicolumn{2}{c|} { $V_{cd}, V_{cs}, V_{cb}$} \\
Angles  & \multicolumn{2}{c|} { $V_{td}, V_{ts}, V_{tb}$} \\     
 &\multicolumn{2}{c|} {}  \\  \hline
 &\multicolumn{2}{c|} {}  \\ 
Lepton Mixing Angles &\multicolumn{2}{c|} {No conventions} \\
 &\multicolumn{2}{c|} {}  \\  \hline
\end{tabular}
\end{center}
\end{table}
 
For each commuting set of generators of
the group, we have an independent coupling, so there are
three gauge couplings $g, g'$, and $g_3$ to be
determined experimentally. 
(I've included $\alpha_s$ because the Yang-Mills Lagrangian can be extended to
include an $SU(3)$ color symmetry to describe QCD).
There are two parameters needed to characterize the Higgs field: 
the vacuum expectation value $v$,  and the
Higgs mass, $M_H = \sqrt{2}\mu^2$.

The experimentally accessible quantities are the coupling constants $G_F,
\alpha_s,$ and $\alpha_{\rm EM}$,  and the gauge boson masses $M_W$ and $M_Z$. The model
parameters and the experimental measurables are easily related by the
following set of equations:

\begin{eqnarray}
M_W^2 & = & \frac{g^2v^2}{4} \\
M_Z^2 & = & \frac{g^2v^2}{2\cos^2\theta_W} \nonumber \\
e & = & g\sin\theta_W  \nonumber \\
G_F & = & \frac{g^2}{8M_W^2} = \frac{1}{2v^2} \nonumber \\
\tan\theta_W & = & \frac{g}{g'}  \nonumber \\
\sin^2\theta_W & = & 1 - \frac{M_W^2}{M_Z^2} \nonumber \\
M_H & = & \sqrt{2}\mu^2 \nonumber
\end{eqnarray}
Very often experimental results are characterized in terms of $\sin^2\theta_W$,
which determines the mixing between the neutral $SU(2)$ and $U(1)$ gauge fields
that
result in the physical photon and the $Z$ boson.

When the matter fields of quarks and leptons are
introduced the number of free parameters
proliferates appallingly.
The fermion-gauge couplings are totally determined by 
$\alpha_{\rm EM}, \alpha_s, G_F$ and 
$M_Z$; however, the fermion masses coming from the 
Yukawa coupling of the fermions to the Higgs are all
free parameters.

We have another set of parameters to introduce in the 
form of a rotation matrix. It
appears that quark flavor eigenstates of strong interactions
 are not eigenstates of the weak interactions and we
need to experimentally determine the $3 \times 3$ mixing matrix that rotates one
basis into the other. This rotation
matrix is called the Cabibbo-Kobayashi-Maskawa (CKM) matrix~\cite{ckm}
and it is thought to
contain the origins of
CP violation.
Finally, if neutrinos have mass (and we have no good reason to think
they 
don't) we have to be
prepared for the neutrinos to mix as well and there is
an equivalent $3 \times 3$ CKM matrix for
lepton sector.

In order for the SM to be completely defined, all these parameters must be
measured. Once the model is defined, we can test it and in fact a major part of
every high energy physics  experiment now and for the
past 20 years has involved testing the predictive power
of the SM.
The depressing but true fact is that so far, in every case, 
either experiment has
confirmed SM predictions or experiment has been wrong!

\subsection{A Little History}

In 1967, Steven Weinberg published a paper~\cite{weinberg}where he
stated: ``Leptons interact only with
photons, and with the intermediate bosons that presumably
 mediate weak interactions.  What
could be more natural than to unite these spin-one
bosons into a multiplet of gauge fields?"  Most of the
ingredients of what would become the SM were in place in the early 1970's~\cite{sm},
and in 1971--72  t'Hooft and Veltman showed that
the theory was renormalizable.~\cite{thooft} 

A stunning feature of the SM was that
it predicted a new interaction: the weak neutral current (NC). 
This was the first time a fundamental interaction was
predicted before it was
 observed. It was  clearly a triumph in 1981 to see $W$'s and $Z$'s
directly, but I personally believe it was 
the observation of the NC that convinced us the SM
was right. 

The
easiest way to look for evidence of
neutral currents is to make $Z$'s directly ($e^+e^- \to Z^0)$ or
($q \bar q \to Z^0$). However,
there was no machine capable of doing that in the early 1970's. The
mass of the
$Z^0$  is approximately 92 GeV. None of the machines available
in the 1970's could produce
92 GeV in the center of mass!

Some of the experimental facilities operating in the early 1970's were:
\begin{itemize}
\item {\bf SLAC:} A linear accelerator that could produce a 22 GeV $e^-$ beam. 
They were also just
turning on an $e^+e^-$ storage ring SPEAR with a
maximum energy of $2.6 \times 2.6$ GeV (5.2 GeV in the center of mass (CM)).
\item {\bf FNAL:} Just turning on with 200 GeV $p$ beam (increased to 400 GeV
by end of the decade). 200 GeV $p$ on a fixed target gives
approximately 20 GeV 
in the CM ($E_{\rm CM} = \sqrt{2E_{lab}m}$) so again they were
 not able to produce $Z$ bosons
directly.
\item {\bf CERN:} A proton
synchrotron produced a 28 GeV $p$ beam which
could be used to make a  $28 \times 28$ GeV $p p$ collider (ISR).
In the
late 1970's, CERN upgraded
the ISR to a $270 \times 270$ GeV  $p \bar p$ 
storage ring which is where the $Z$ was
directly
produced and detected for the first time.
\item {\bf BNL} A 33 GeV $p$ beam on fixed target.
\end{itemize}

\subsection{The Discovery of Neutral Currents}

\begin{figure}
\centerline{\psfig{figure=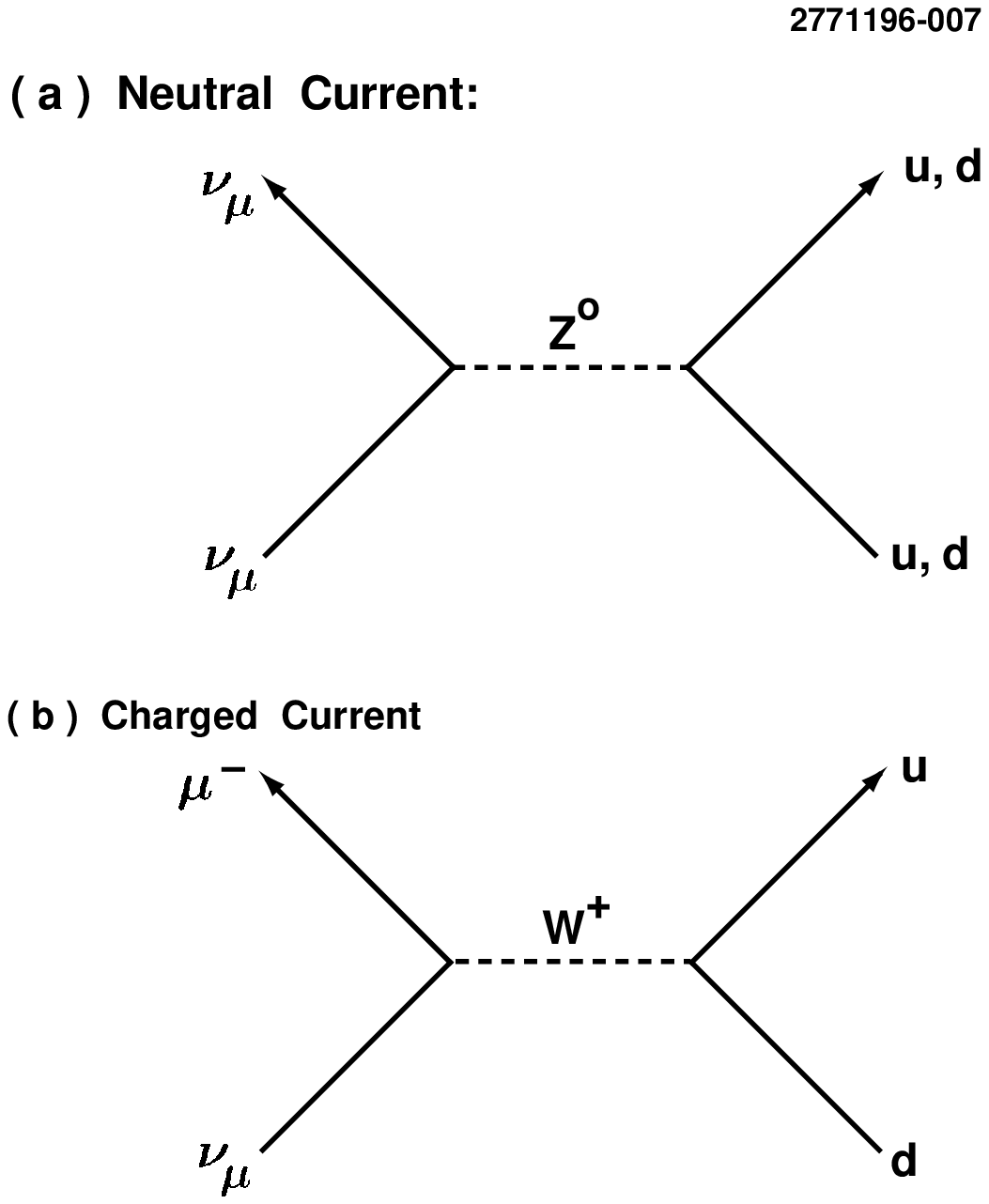,height=3.0in}}
%\vskip 3.0in
\caption{Diagrams for (a) neutral current and (b) charged current neutrino scattering.
\label{fig:ncnu}}
\end{figure}

Because no machine could produce $Z$ bosons directly in the early 1970's,
the first 
observation of NC had to be indirect, and the most sensitive technique
was neutrino scattering, 
 where a
$\nu_\mu$ scattered off the quarks in a nuclear target
as shown in Figure ~\ref{fig:ncnu}. The hadrons in
the final state could be detected (the incoming and outgoing $\nu$ were
invisible to the detector), and the absence of a muon meant it was a NC event.
The rate for the NC process could be compared to the rate for the
corresponding charged current (CC) process where in addition to the hadrons from
the breakup of the nucleon, an accompanying muon could be seen.

It is straightforward to work out the cross sections for $\nu$ and $ \bar \nu$
scattering off nucleons. They are discussed in detail in Quigg~\cite{quigg} if you are
interested. 
The ratios of cross sections are what can be experimentally measured most
precisely and the experimentally accessible quantities are:
\begin{eqnarray}
R_\nu \equiv \frac{\sigma(\nu N \to \nu X)}{\sigma(\nu N \to \mu^- X)}
& = &\frac{1}{2} - \sin^2\theta_W + \frac{20}{27}\sin^4\theta_W \\
R_{\bar{\nu}} \equiv \frac{\sigma(\bar{\nu} N \to \bar{\nu} X)}
{\sigma(\bar{\nu} N \to \mu^+ X)} & = & \frac{1}{2} - \sin^2\theta_W + 
\frac{20}{9}\sin^4\theta_W 
\end{eqnarray}
Measuring absolute cross sections involves a detailed knowledge of the flux
of the incident $\nu$ beam and that is hard to know; however, in the ratio, both 
the flux and
the poorly known energy spectrum of the $\nu$ beam cancel.

By just seeing $\nu N \to \nu X$ reactions, one observes NC for the first time
which is a great achievement; however, one can also use the cross section
ratio to extract $\sin^2\theta_W$ (or whatever your favorite third
$SU(2) \times U(1)$ parameter is; the convention was to use $\sin^2\theta_W$
until LEP came on line and now $M_Z$ is standard).
Once $\sin^2\theta_W$ is known, all the SM couplings are determined and
by measuring
$R_\nu$ and $ R_{\bar \nu}$ one gets a wonderful consistency check. 
If both give the same
value of $\sin^2\theta_W$, it is an indication one has chosen the right gauge
structure (for example, $SU(2)_L \times SU(2)_{R} \times U(1)$ would predict different
relations between $R_\nu$ and $R_{\bar \nu}$) and one
is  starting to test the predictive
power of the theory.

To make neutrinos, one starts with a proton beam on a target that  produces
lots of secondary particles; in particular, lots of kaons and pions will be
produced.  The kaons and pions are selected for sign and then allowed to decay,
($\pi \to \mu \nu_\mu, K\to  \mu \nu_\mu, 
 K\to \pi \mu \nu_\mu$,) and neutrinos are produced.  Muon
 neutrinos are strongly favored by helicity,
and neutrinos or antineutrinos are selected by the charge of the meson.
The experiments are hard. The major obstacle is just rate. The $\nu N$
scattering cross section is proportional to $G^2_F M_p E_\nu$ and $G_F$ is a small
number so the cross section is small.
\begin{eqnarray}
\sigma^{\nu N}_{CC} &\sim& 6 \times 10^{-6} {\rm nb} (E_\nu/{\rm GeV)}/{\rm nucleon}
\end{eqnarray}
Working at the highest possible $\nu$ beam energy is clearly an advantage.
FNAL with its 200 GeV $p$ beam 
had a great advantage for making high energy neutrinos over CERN with 30 GeV $p$, 
but
CERN got there first.

\begin{figure}
%\vskip 4.5in 
\centerline{\psfig{figure=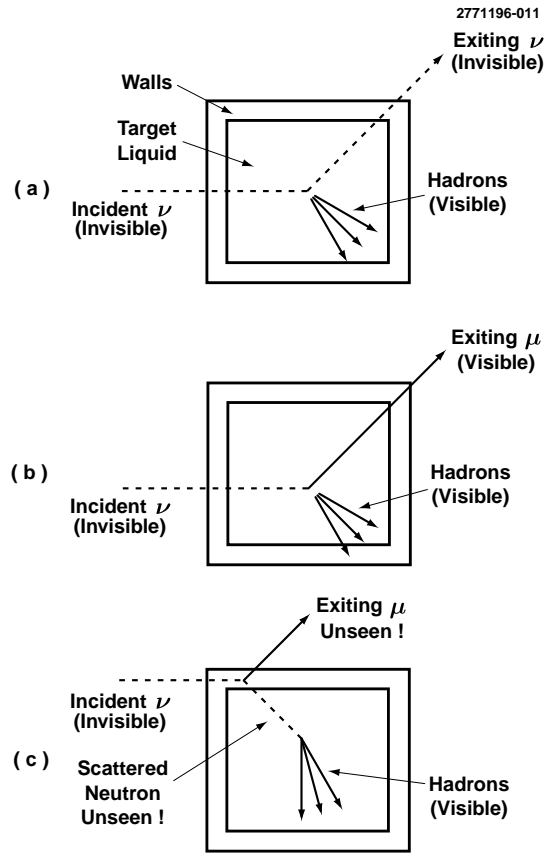,height=4.5in}}
\caption{The signature of a NC event is illustrated in (a) where the
incoming and escaping neutrinos are invisible and the signal is the observation
of a hadronic cluster.  A CC event is shown in (b) where the exiting muon is
observed.  A background event is shown in (c) where an incoming neutrino interacts
in the material of the chamber wall, producing a neutral hadron which cannot be
detected and a muon that escapes.  The neutral hadron can then interact in the
chamber and look like a NC event. 
\label{fig:gargamelle}}
\end{figure}

The detector that made the discovery was the 12' Gargamelle bubble 
chamber~\cite{gargamelle}.  The central part of the detector was a big tank of 
supersaturated
freon.  When a charged particle passed through the supersaturated gas, it left an
ionization trail.  The gas was expanded suddenly after the beam pulse and bubbles
formed along the ionization trail.  The bubble tracks were then photographed,
scanned, and measured by hand for evidence of interesting physics processes.
An important feature of the detector was
the ability to identify NC events by having good
solid angle coverage for muons so a muon could not escape undetected.
Figure ~\ref{fig:gargamelle} illustrates what CC, NC and background events would
look like in the detector.  The most worrisome background
was a $\nu$ interacting in the material of the chamber wall, producing a
neutral hadron and an escaping muon. The neutral hadron could
then interact in the chamber and look like NC event.

The experiment took
some 300,000 pictures, 83,000 with the $\nu$ beam
and 207,000 with the $\bar{\nu}$ beam.  They collected
twice as many events for the $\bar\nu$ beam since the scattering cross section was
approximately one third the $\nu$ cross section due to helicity effects.

\begin{figure}
%\vskip 4.0in
\centerline{\psfig{figure=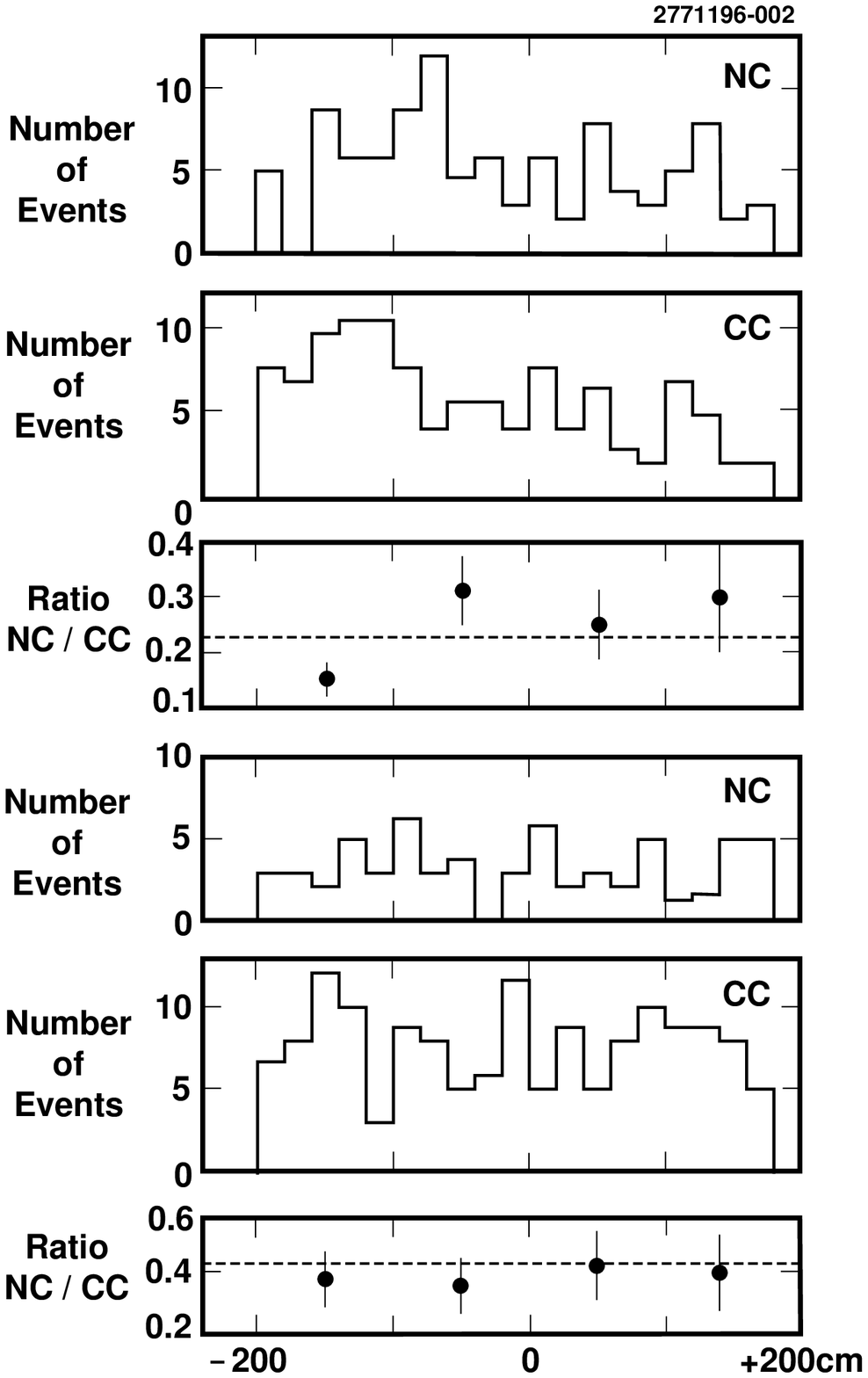,height=4.0in}}
\caption{The number of NC and CC candidate events from the Gargamelle neutrino
scattering experiment as a function of length along the chamber.  The top three
plots show the data for the $\nu$ beam and the bottom three plots show the
$\bar\nu$ beam data. 
\label{fig:gdata}}
\end{figure}

The experimenters spent a 
 great deal of effort studying possible backgrounds to the NC sample from
neutral hadrons. One particularly convincing check was that background from
neutral hadrons was expected to show  attenuation along the length of
the chamber and they were able to show that their NC candidates
 had a uniform distribution along the chamber length as shown in Figure
~\ref{fig:gdata}.

From the ratio of the number of NC to CC events, they
were able to conclude  ``$\sin^2\theta_W$ is in the range 0.3--0.4.''
They conservatively claimed ``if the events are due to NC, then 
$R_\nu$ and $R_{\bar \nu}$ are
compatible
with the same value of $\sin^2\theta_W$.''

This experiment has been repeated many times since 1973.
The best experiment to date was done using a 450 GeV proton beam
and the CDHS detector and was published in
1990~\cite{cdhs}.  It gives, using similar techniques:
 \begin{equation}
\sin^2\theta_W = 0.228\pm 0.013 (m_c -1.5) \pm 0.005 ({\rm experimental}) \pm
0.003 ({\rm theoretical})
\end{equation}
where $m_c$ is the charm quark mass.  

\subsection{The Discovery of Neutrinoless Neutral Currents}

\begin{figure}
%\vskip 2.75 in
\centerline{\psfig{figure=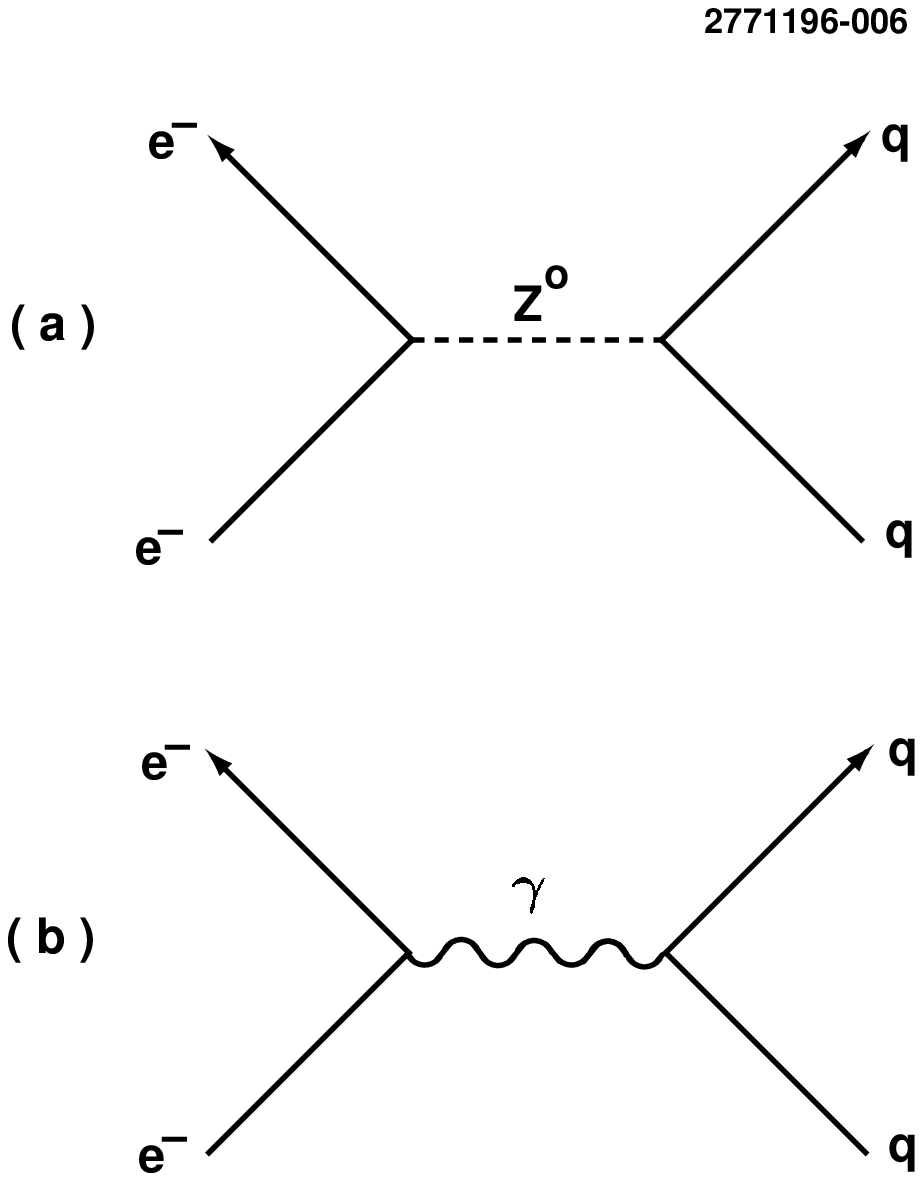,height=2.75in}}
\caption{The two processes, (a) $Z$ exchange and (b) $\gamma$ exchange, which can
contribute to electron-quark neutral current scattering.
\label{fig:zgam}}
\end{figure}

Neutral currents were discovered by neutrino scattering experiments. The
coupling that was observed was consistent
with the SM predictions; however, there are many other terms in the SM Lagrangian
 involving NC
apart from neutrino-quark interactions. In particular, there are NC terms in the
Lagrangian that do not
involve neutrinos (e.g., electron-quark scattering through $Z$ exchange).
These terms are particularly interesting because electron-quark scattering can take
place via $Z$ or $\gamma$ exchange 
as shown in Figure~\ref{fig:zgam}
and the two processes can interfere. As I'll
show, this interference allows one to explore the parity
violating nature of the NC interaction.

There were two approaches that experimentalists used to probe the
electron-quark coupling. The first approach was to scatter
high-energy polarized electrons off of a nuclear target. This was first done at SLAC and
I'll talk about it in some detail. The other was to use atoms. The $e^-$
in an atom interacts with the nucleus
 both via the usual EM interaction and by $Z$ exchange. The
immediate consequence is that the atomic Hamiltonian does not conserve parity.
Everything you learned about stationary states of atomic systems being
eigenstates of the parity operator was incorrect
(although a good approximation)!

In the late 1970's, experiments in atomic bismuth {\it failed} to detect
parity violation, in contradiction with the SM expectation ~\cite{bismuth}. The early atomic
physics results were wrong. Later experiments in the early 1980's 
 agreed with the SM predictions~\cite{thallium} but by
that time the SLAC experiment had already confirmed the SM predictions for
electron-quark couplings in 1978--79.

The basic idea of the electron-quark scattering experiments is that the scattering cross section
is the square of the sum of the weak and EM amplitudes, $A_{WK}$ and $A_{EM}$, 
which can interfere:
\begin{equation}
\sigma \sim |A_{EM} + A_{WK}|^2 = A_{EM}^2
 \left( 1 + \frac{2A_{EM}A_{WK}}{A_{EM}^2} + \frac{A_{WK}^2}{A_{EM}^2}\right)
\end{equation}
At low $Q^2$ ($Q^2 < 10$ GeV$^2$), 
$A_{EM}\gg A_{WK}$ and the last term can be dropped.  The interference term, however,
can be detected.

If we treat the NC as current-current interaction with vector ($V$) and
axial vector ($A$) parts (recall the CC
is $V-A$ but the NC is much more complicated) then
\begin{equation}
A_{WK} = J_e J_q = (V_e V_q + A_e A_q) + (V_e A_q + A_e V_q)
\end{equation}
where the subscripts $e$ and $q$ refer to the electron and quark currents.
The first term is a scalar ( $(V_e V_q + A_e A_q) = A_{WK,{\rm scalar}}$)
and is extraordinarily difficult to detect.  The second term,  ($(V_e A_q +
A_e V_q) = A_{WK,{\rm pseudoscalar}}$) however, is a pseudoscalar
 and has a very nice signature because it changes sign under parity
transformation.
It is straightforward to show that if we define the asymmetry, $\delta$, as the
difference in the scattering cross section for left and right handed scattering
divided by the sum, then:
\begin{eqnarray}
\delta  & = & {\sigma_R - \sigma_L \over \sigma_R + \sigma_L} \\
 & = &{2(A_{WK,{\rm pseudoscalar}}A_{\rm EM})  \over A_{EM}^2}
\end{eqnarray}
where $\sigma_R, \sigma_L$ are the cross sections for right and left handed
coordinate systems and the handedness of the coordinate system is determined by,
for example, 
the longitudinal
polarization of the incoming $e^-$ beam.

The asymmetry is small!  At $Q^2 \sim 10$ GeV$^2$, the ratio of the weak and 
electromagnetic amplitudes can be estimated:
\begin{eqnarray}
A_{EM} & \sim &{4\pi\alpha_{EM} \over q^2} \\
 A_{\rm WK}  &\sim & G_F \\
\delta & \sim & {G_F q^2 \over 4\pi\alpha_{EM}} \\
 & \sim &10^{-4}
\end{eqnarray}
In the SLAC experiment that discovered neutrinoless NC,
high-energy polarized electrons were 
scattered off of an unpolarized  deuterium target~\cite{prescott}.
The scattered  electrons were detected
at a fixed scattering angle in the lab which corresponds to a fixed energy
of the scattered electron.
It is straightforward (but tedious) to start from the SM Lagrangian
 and calculate the expression
for the asymmetry in scattering left versus right handed electrons~\cite{quigg}.

To measure an asymmetry of $10^{-4}$ to 10\% precision, one needs
$10^{10}$ events. Clearly one cannot count scattered electrons one by one. 
The experiment used a slightly
different philosophy from the usual single particle counting techniques common in
high energy physics.  Instead of counting the
scattered electrons individually, the detector integrated the signal
and measured a  current of scattered electrons on each beam
pulse.

\begin{figure}
%\vskip 1.7in
\centerline{\psfig{figure=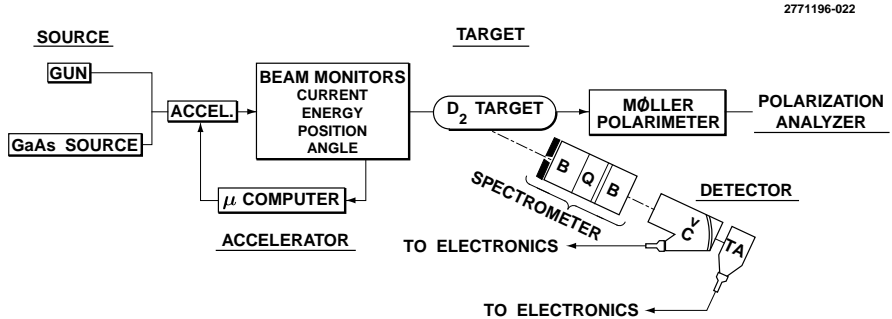,height=1.7in}}
\caption{Overview of the SLAC polarized electron scattering experiment that
discovered neutrinoless NC.
\label{fig:slacnc}}
\end{figure}

\begin{figure}
%\vskip 2.75in
\centerline{\psfig{figure=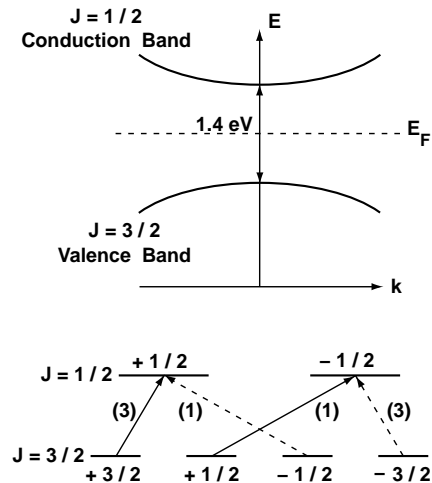,height=2.75in}}
\caption{The energy levels of the conduction band and valence band of GaAs.
A circularly polarized laser which induces transitions can polarize the upper
state as shown.
\label{fig:gaas}}
\end{figure}

Figure ~\ref{fig:slacnc} shows
an overview of the experiment. At the gun end of the
linac, they started with a polarized  $e^-$ source.
The polarized $e^-$ source was really very cute. Usually in a linear accelerator,
one  uses a
thermionic cathode which is heated up and electrons are boiled off, collected, 
and used to produce an unpolarized beam.
To make polarized electrons, they replaced the thermionic cathode with a gallium arsenide
crystal.
The electrons were polarized by optically pumping electrons 
from the $j=3/2$ valance band to
$j=1/2$ conduction band of the crystal with a circularly
polarized laser beam (710 nm light).
Starting from the valance band a circularly polarized
 photon has $\Delta j_Z = +1 ~{\rm or} -1$. The Clebsch-Gordon 
coefficients are favorable and one gets $3$ times as many electrons in one $m_j$ level as
the other in the upper state conduction band,  which polarizes the upper state.
This is illustrated schematically in Figure ~\ref{fig:gaas}.

To get the electrons  out of the crystal conduction band, they coated
the  surface with cesium and oxygen
which produced a negative work function.
The  electrons could  escape and their polarization was preserved.
The circular polarization of the laser controlled the  polarization of the 
$e^-$ beam and it could be changed on a pulse by pulse
basis in a random way.  This technique theoretically could produce an electron
beam with 50\% polarization.  In practice, the average electron beam polarization
was 37\%.  

The beam was then accelerated down
the linac with very little loss of polarization.  At the
end of the linac, the beam was deflected into the  beam switchyard
onto the deuterium target.

The scattered $e^-$ flux was measured with 2 independent detectors, both measuring
the total
charge passing  through them.
The polarization of the spent beam was determined with a M\"oller polarimeter, taking
advantage of the asymmetry in the cross section for a longitudinally polarized electron
scattering on polarized target electrons.
The parity violating asymmetry that was
the signature for NC was computed by
counting 
electrons scattered into the detector
when the electron beam was left handed versus right handed.

The challenge of this type of experiment is not just to measure an asymmetry of one
part in $10^4$, but to convince yourself you are measuring the correct
asymmetry!  A great deal of attention was paid to ensure that all possible instrumental
asymmetries were at the $10^{-5}$ level or smaller.
The final result was 
\begin{eqnarray}
\sin^2\theta_W &  = & 0.222\pm0.018.
\end{eqnarray}

It was a demonstration of the power of the 
SM  that with a single value of the parameter
$\sin^2\theta_W$, it could account in detail for the strengths of  very
disparate processes: both the SLAC polarized electron scattering experiment
and the neutrino scattering experiments.
And that is why, with the SLAC result, the high energy
community was convinced, even before
the $Z$ was found, that the SM was correct and that $SU(2)_L \times U(1)$ was the
correct gauge group to describe the world around us.

\section{Anatomy of an Experiment}

\subsection{Overview}

Having talked about the experiments done in the 1970's that discovered NC, I now
want to fast forward to present day. I will start by describing the landscape of
experimental high energy physics today. What are the current machines and
detectors? Where is the physics happening?

\begin{figure}
%\vskip 5.5in
\centerline{\psfig{figure=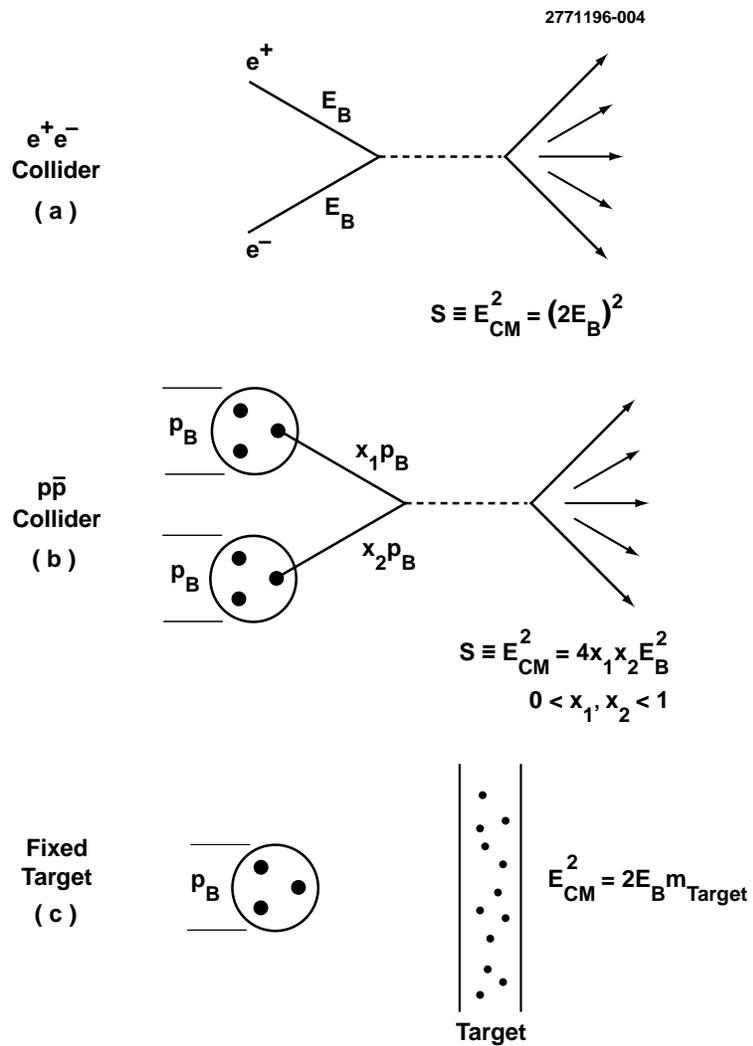,height=5.5in}}
\caption{The basic types of collisions are illustrated.
\label{fig:overview}}
\end{figure}

There are 3 basic types of machines currently operating: $e^+e^-$ colliders,
a $p\bar p$ collider, and fixed target experiments. The $e^+e^-$ colliders are
machines where bunches of $e^+$ and $e^-$ with equal energy and opposite momenta
collide. All of the beam energy is available in the center of
mass(CM) as illustrated in Figure~\ref{fig:overview}(a), and the CM is the lab frame.
The dominant process is annihilation. Rates at these machines tend to be low
because the annihilation cross section is small and falls with increasing energy.

At the Fermilab (FNAL) $p\bar p$ collider, often called
the Tevatron, bunches of $p$ and $\bar p$ collide with equal
and opposite momenta as shown
in Figure~\ref{fig:overview}(b). 
The partons in the protons that interact carry only about
1/6 of the energy of the incident proton, although the distribution of the
fraction of energy carried by the partons has a long tail. The CM energy of
the parton-parton interaction is not known event by event, and, in fact,
the CM frame of the interaction and the lab frame may not be the same.  The CM may
have appreciable boost along the beam direction in the lab. The total inelastic
cross section is very large, which results in large backgrounds to the signals one
wants to see. There are lots of events and sorting between interesting and
uninteresting events is a challenge.

Fixed target experiments usually involve production of a secondary beam of
particles to be studied by slamming protons into a target. Examples are production
of kaon beams to study rare $K$ decays or CP violation in the $K$
system, or production of $\nu$ beams to study $\nu$ oscillations or for deep inelastic
scattering experiments.   Here, the CM energy available is only a fraction of
the beam energy as illustrated in  Figure~\ref{fig:overview}(c).

Table ~\ref{tab:summary} shows the kind of physics accessible with the major high energy
physics facilities in the world by type of collision and CM energy.
I have not tried to be exhaustive and only listed the main players at each machine,
excluding a host of smaller experiments, especially in the fixed target program. I
will mostly talk about the CDF, ALEPH, and CLEO experiments.

\begin{table}[t]
\caption{The major experiments currently operating.
\label{tab:summary}}
\vspace{0.4cm}
\begin{center}
\begin{tabular}{|c|c|c|c|c|c|}
\hline
 & & & & &  \\
Machine&Lab&Detector&Type of&$\sqrt{s}$&Physics\\
						&					&       &Collision& (GeV) &         \\
      & & & & &  \\
 & & & & &  \\ \hline
 & & & & &  \\
Tevatron & FNAL & CDF & $p \bar p$& 1800 & $t, W, Z$  \\
& & D0 & & & \\ \hline
HERA & DESY & H1 & $ep$ & $300 \times 820$ & QCD, exotics \\
& & ZEUS & & & \\ \hline
LEPI(II)& CERN & ALEPH & $e^+e^-$ & 92(195) & $Z(W$,Higgs) \\
& & L3 & & &  \\
& & OPAL & & & \\
& & DELPHI & & & \\ \hline
SLC & SLAC & SLD &  $e^+e^-$ & 92 & $Z$ \\ \hline
CESR & CORNELL & CLEO &  $e^+e^-$ & 10.58 & $B$ \\ \hline
BEPC & CHINA & BES &  $e^+e^-$ & 4 & $\tau,\psi$ \\ \hline
AGS & BNL &  \multicolumn{4}{c|} {30 GeV $p$ $\to K$ beams } \\ \hline 
LAMPF & LANL &  \multicolumn{4}{c|} {$\bar \nu_\mu$ beams} \\ \hline
Tevatron & FNAL &  \multicolumn{4}{c|} {1TeV $p$ $\to \nu,K$ beams} \\ \hline
SPS & CERN &  \multicolumn{4}{c|} {450GeV $p$ $\to \nu,K$ beams} \\ \hline
\end{tabular} 
\end{center}
\end{table}

\subsection{Detectors}

When the beams collide at an accelerator, physics happens: particles that we
want to study emerge. The interaction region is instrumented with a detector that is
designed to record as much information as possible about what is emerging from
the beam collision.

The form of the detector depends in its gross geometry on the accelerator type.
At storage rings where the lab frame is also the CM frame for the interaction,
outgoing particles from the
interaction are nearly isotropically distributed about the collision point
and detectors reflect that fact. The detectors try to surround as much of
the solid angle around the interaction point as possible, given practical and
financial constraints. Typically such detectors are forward$-$backward and
azimuthally symmetric to reflect the production symmetry and cover over
$90\%\times 4\pi$ of the solid angle.  A typical collider detector is shown
in Figure~\ref{fig:alephdet}

\begin{figure}
%\vskip 3.35in
%\centerline{\psfig{figure=aleph.ps,height=3.35in}}
\centerline{Picture converted to gif file}
\caption{A schematic view of the ALEPH detector which operates at the
LEP collider. (http:\-//alephwww\-.cern\-.ch\-/\-alephgif\-/\-alephpict\-.html)
\label{fig:alephdet}}
\end{figure}

\begin{figure}
%\vskip 4.5in
\centerline{\psfig{figure=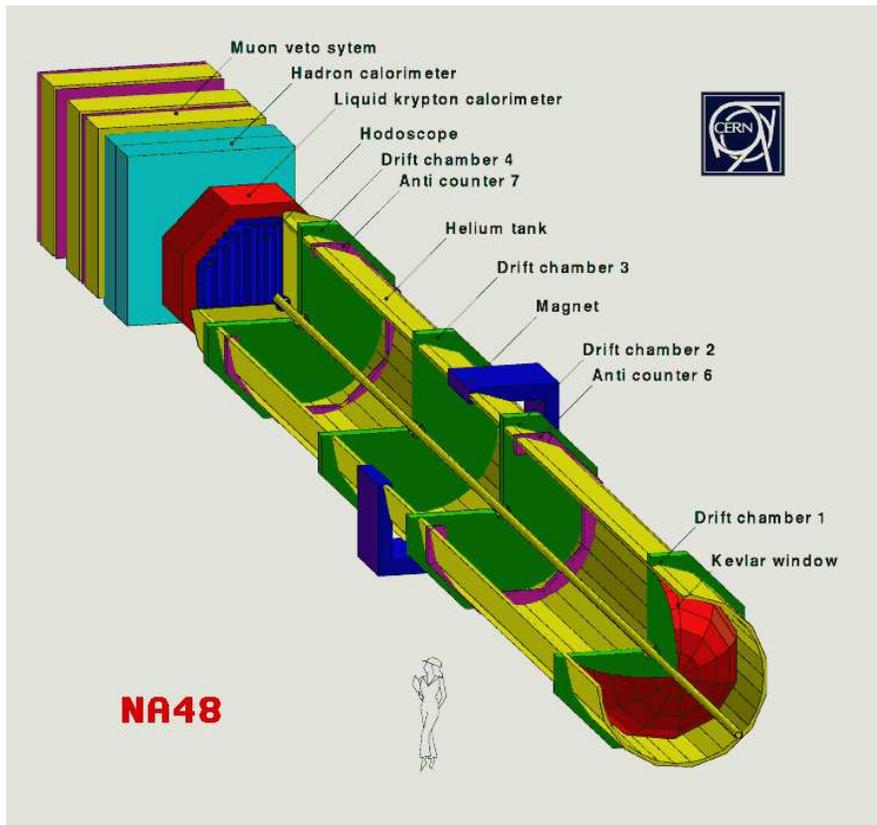,height=4.25in}}
\caption{A schematic view of the NA48 detector which is a fixed target 
experiment operating the the kaon beam at CERN. (http:\-//\-www1.\-cern.\-ch\-/NA48\-/\-Wel\-come\-/\-im\-a\-ges\-/\-de\-tec\-tor\-.\-html)
\label{fig:na48}}
\end{figure}

In a fixed target experiment, the interactions are very boosted. The experiments
can cover most of the solid angle in the CM frame by being very long and narrow
in the lab frame as shown in Figure~\ref{fig:na48}.

It is not possible to describe a generic detector.  Each detector
is individually  designed to
match the machine at which it runs; however, all detectors are composed from a
fairly consistent set of building blocks which can be easily described, although
the execution or techniques used on different experiments will vary widely.

The basic components of all detectors are:
\begin{itemize}
\item charged particle tracking which determines the momentum and charge of
charged tracks
\item electromagnetic (EM) calorimetry which identifies photons and electrons and measures their
energy and direction.
\item hadron calorimetry which is used to measure the energy of jets of hadrons
\item muon detection which is used to identify muons
\item particle identification of various sorts to distinguish different types
of hadrons, particularly pions and kaons.
\end{itemize}
I will briefly discuss the various detector elements and how they are
most commonly used.~\cite{klein}

\subsubsection{Charged Particle Tracker}

\begin{figure}
%\vskip 5.5in
\centerline{\psfig{figure=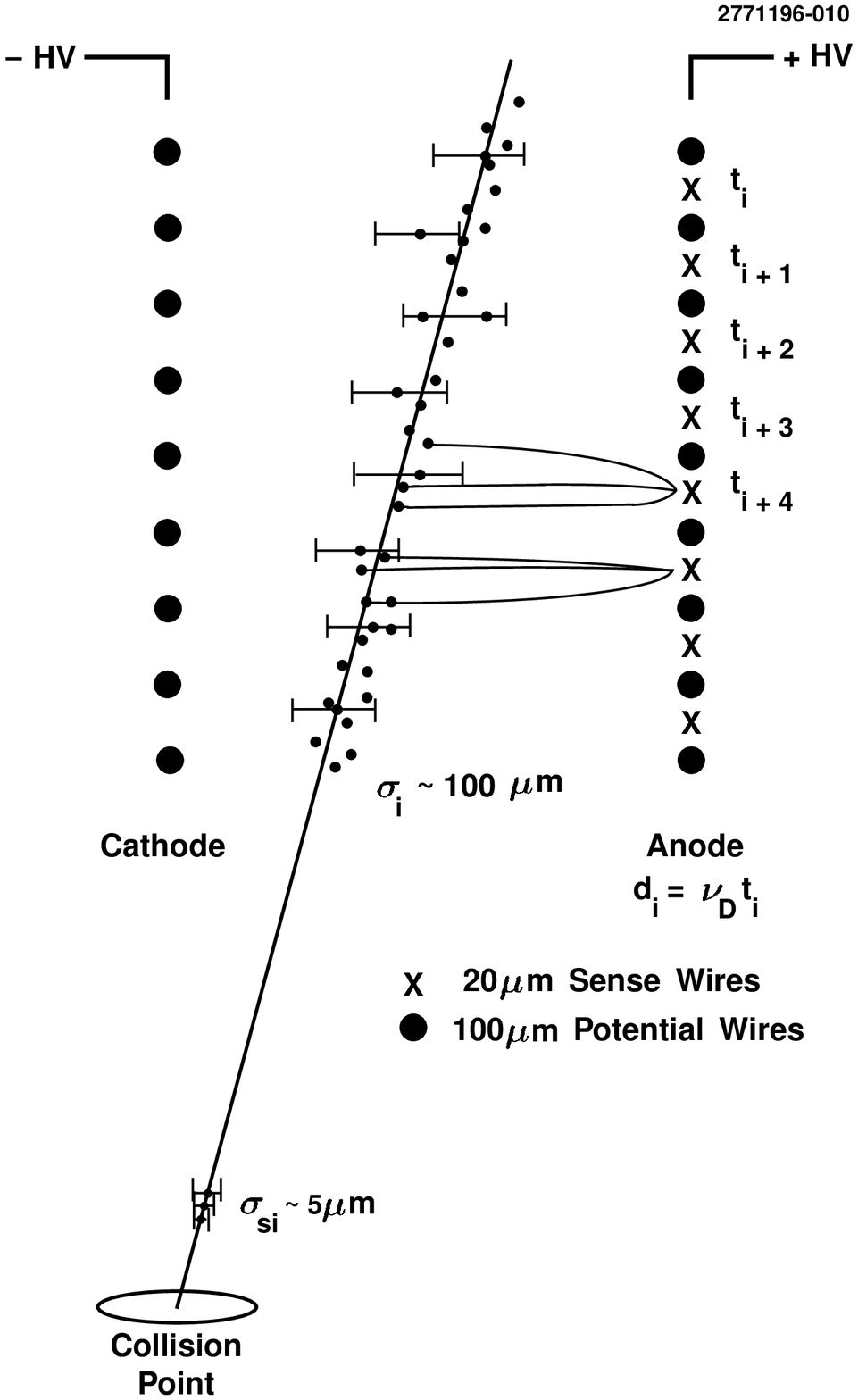,height=5.5in}}
\caption{A schematic view of how a drift chamber operates.  An incoming charged
particle ionizes the gas.  The liberated electrons drift in the applied electric
field to the anode wire where a signal is recorded.
\label{fig:dc}}
\end{figure}

A charged particle tracker, usually a drift
chamber, is at the heart of most experiments.
A basic drift chamber is made of cathode wires at negative high voltage (-HV),
 and anode wires at positive high voltage (+HV), enclosed 
in a gas
volume.
Incoming charged particles passing through the gas ionize the atoms
in the gas. In the
ionizing encounter, electrons are liberated and drift in the applied electric
field towards
the anode as shown in Figure~\ref{fig:dc}.
To measure the position of a track, a clock is started when the
particle is produced (at the
beam crossing) and stopped when the pulse height on a wire 
exceeds a preset value.
Associated with each wire there will then
be a time $t_i$. Using $d_i = v_D t_i$ where
$v_D$ is the drift velocity of electrons in the gas, one can infer the
distance from the wire to  where the track ionization segment came from.
By joining hits, one defines the track of the incident charged particle.

Precision silicon tracking devices work on the same physics principle, although the
anode and
cathode in a silicon detector are no longer wires but
electrodes etched on a  thin silicon wafer.  Silicon detectors are
usually placed right around the beam pipe and provide high resolution
 position measurements on
tracks close to the interaction point.

The entire tracking volume is usually enclosed in a uniform magnetic field and from
the curvature of tracks one measures the particle's momentum.

\begin{figure}
%\vskip 3.5in
\centerline{\psfig{figure=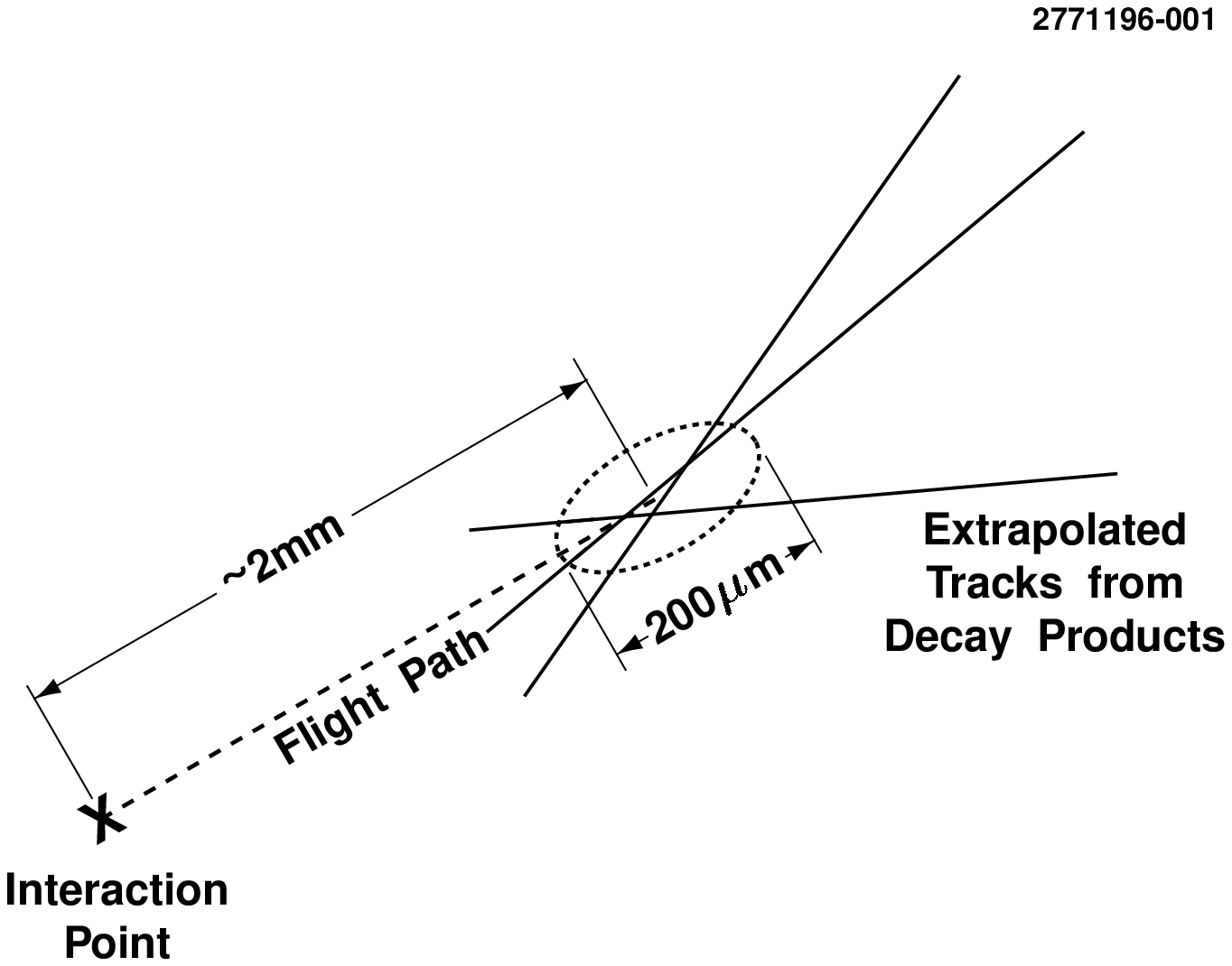,height=3.5in}}
\caption{A sketch of a typical $B$ meson decay as seen by the
LEP collider experiments.  The mean flight path of a $B$ meson at LEP
is approximately 2mm, and the vertex resolution of the silicon detectors
is about 200 $\mu$m.
\label{fig:displ}}
\end{figure}

\begin{figure}
%\vskip 5.0in
\centerline{\psfig{figure=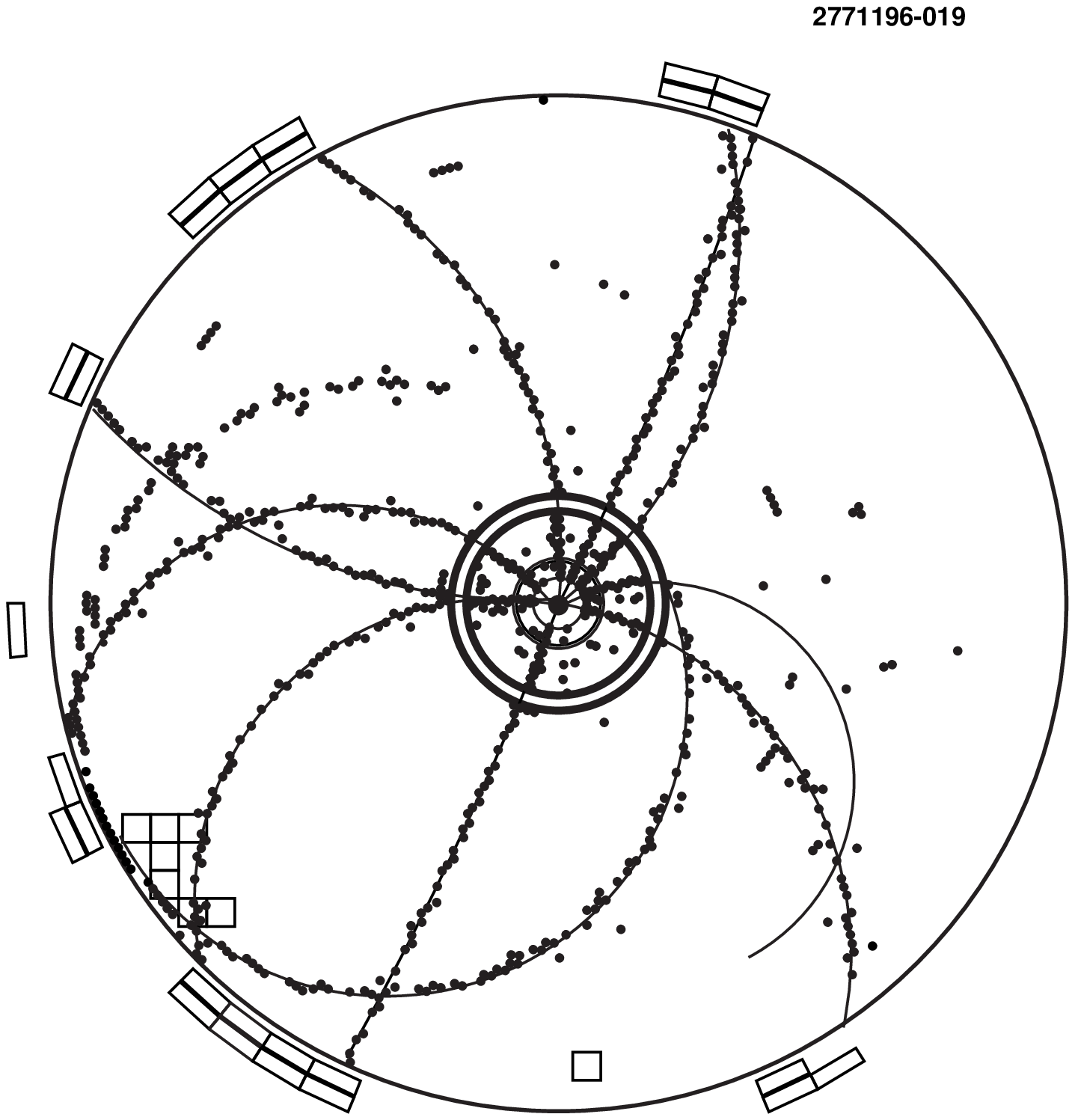,height=5.0in}}
\caption{A typical event in the CLEO tracking chambers.  The dots are hits
on anode wires and the lines are drawn by the pattern recognition software
which is responsible for joining the dots into tracks.
\label{fig:track}}
\end{figure}

Drift chambers are the most versatile of all detector elements. In addition to
measuring momentum and charge, tracks left by charged particles can be
extrapolated back to the interaction point. Tracks with significant impact
parameters to the beam crossing point, or 
that can be combined to form a
displaced vertex as illustrated in Figure~\ref{fig:displ},
may come from the decays of long-lived particles.
For example, the silicon vertex resolution for the LEP experiments is $\sim 200\mu$m
while the typical decay lengths of heavy flavor $(\tau, D, B)$ particles
are $ \sim 2$ mm.
Figure~\ref{fig:track} shows tracks in a typical collider detector.  The dots are hits on 
anode wires.  The pattern recognition software is responsible for joining the
dots into tracks.

\subsubsection{Electromagnetic Calorimeter}

Most experiments have some form of electromagnetic (EM) calorimeter.
EM calorimeters are devices
where electrons  and photons will shower in an  alternating
 sequence of bremsstrahlung
and pair production, giving up all their energy.
This is the primary form of photon detection. Photons deposit all
their energy in the EM calorimeter, and they are identified as photons
(as opposed to electrons which will also shower)
because there is no charged track pointing
at the cluster of energy.  The photons can then be combined with other photons to
reconstruct $\pi^0$'s from their decay $\pi^0 \to \gamma\gamma$.

\begin{figure}
%\vskip 3.5in
\centerline{\psfig{figure=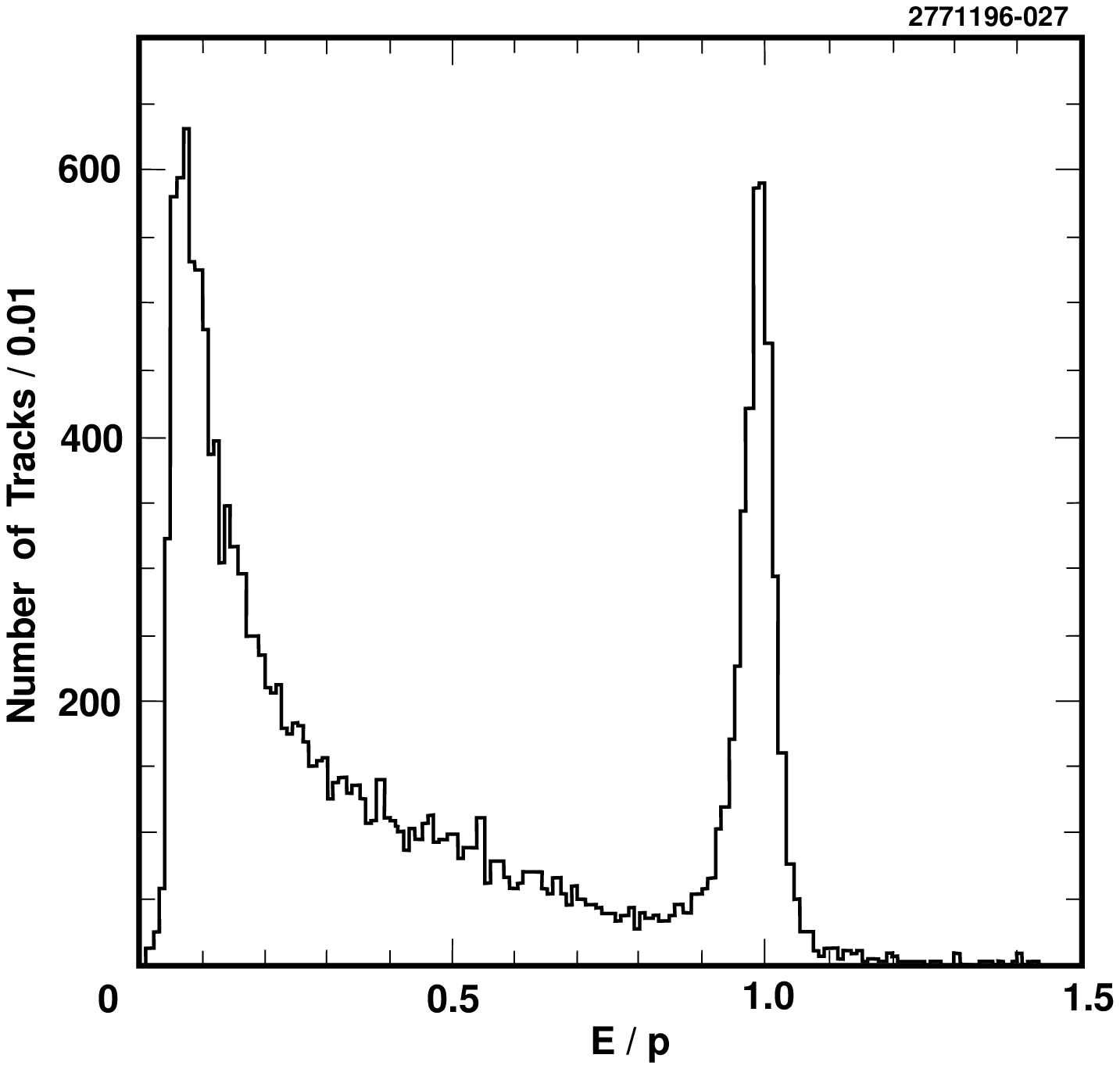,height=3.3in}}
\caption{A plot of the ratio of energy (E) in a cluster in the CLEO
EM calorimeter
to the momentum (p) of a track pointing to that cluster.  The peak at 1 is
due to electrons.  The clusters with E/p less than 1 are mostly hadrons.  
\label{fig:eid}}
\end{figure}

EM calorimeters are
also very powerful as $e^-$ detectors. An electron
is identified by matching the energy of a shower in the calorimeter to the
momentum measured on a charged track pointing to the
cluster.  Electrons are easily separated from hadrons and muons,
which deposit much less energy as shown in Figure~\ref{fig:eid}.

EM calorimeters are built using a variety of techniques. The most precise are the
 crystal
calorimeters which are constructed of
blocks of  CsI, NaI, or BGO.  The entire EM shower is 
contained in the uniform crystal blocks with dimensions typically
5 cm $\times$ 5 cm $\times$ 30 cm deep. Adjacent blocks are summed to
reconstruct the shower.

An alternate technique is
a sampling shower detector typically made of a lead-scintillator sandwich.  Thin lead
plates are alternated with
a scintillator such as liquid argon.  The shower forms in the
lead and is then sampled in the scintillator.
EM calorimeters can be small. The CLEO CsI calorimeter with a depth of 30 cm can
 fully
contain a 5 GeV photon
with little light leakage.  The ALEPH Pb-liquid argon EM calorimeter
identifies electrons with tens of GeV of energy and is only 40 cm in radial depth.

\subsubsection{Hadron Calorimeters}

Hadron calorimeters, in contrast to EM
calorimeters, are big. When a strongly interacting particle goes through
material, there are elastic and inelastic interactions with nuclei in the material,
producing secondary hadrons. Hadron calorimeters typically use a sampling
technique with plates of a dense high $Z$ material such as uranium or iron
sandwiching a scintillating material or ionization detector where the shower is
sampled.
Again, the idea is to get a particle to give up all its energy in the
calorimeter.
Typical hadronic interaction lengths of materials such as iron are 15-20cm, and
many interaction lengths are needed for an efficient detector. 
Since the hadron calorimeter in a colliding beam detector has to go outside the
drift chamber and EM calorimeter, this can be a lot of iron!  The ALEPH
hadron calorimeter is 1.2m thick, starting at a radius of 3m from the
beam line.

The most important use of a hadron calorimeter
is to measure the energy of dense jets of particles. In CDF, the
energy of the jet is used to infer the energy of the underlying
parton that produced the jet, and we will come  back to this when I
talk about the measurement of the top quark mass.    

An important use of both EM and hadron calorimeters is to detect neutrinos
in an event.  Neutrinos will leave no
measurable signal in the detector, so  the only hope is to detect 
them indirectly. This is particularly important for the $W$
and top quark discoveries and mass measurements at a hadron machine.
The experiments use a missing momentum technique. I
 mentioned that at a hadron machine, the CM of the parton-parton collision 
is not necessarily 
the lab frame.
Since fragments of the parent proton and antiproton escape down
the beam pipe in the very forward direction, there is no way to use
conservation of total momentum in the event to infer the
momentum of the unobserved neutrino.  However, the components
of the momentum in the plane transverse to the
beam line ($p_T$) can be measured for all the 
observed
decay products by using the vector sum over the
energy deposited in the calorimeters, and
that should be zero before and after the collision.
Therefore, the neutrino transverse momentum can be
inferred as the negative of the vector sum of all
the transverse momenta detected  in the event.

\subsubsection{Muon Detectors}

Muons are very penetrating and so muon detectors are typically
planar drift chambers outside of the calorimeters and the magnet flux return.
Any charged particle that makes it through that many interaction lengths of
material is
identified as a muon.

\subsubsection{Particle Identification}
 I
have already talked about how to identify electrons, muons
and photons. Many experiments
find it useful to also distinguish protons, pions and kaons. There are currently
experiments with very sophisticated particle identification
 systems based on differences in the
pattern of Cerenkov radiation emitted by the various particle species.
Low energy experiments can get some information
from time of flight or ionization losses in their drift chambers, but the information
is limited. 

\subsubsection{Examples}
\begin{figure}
%\vskip 5.0in
\centerline{\psfig{figure=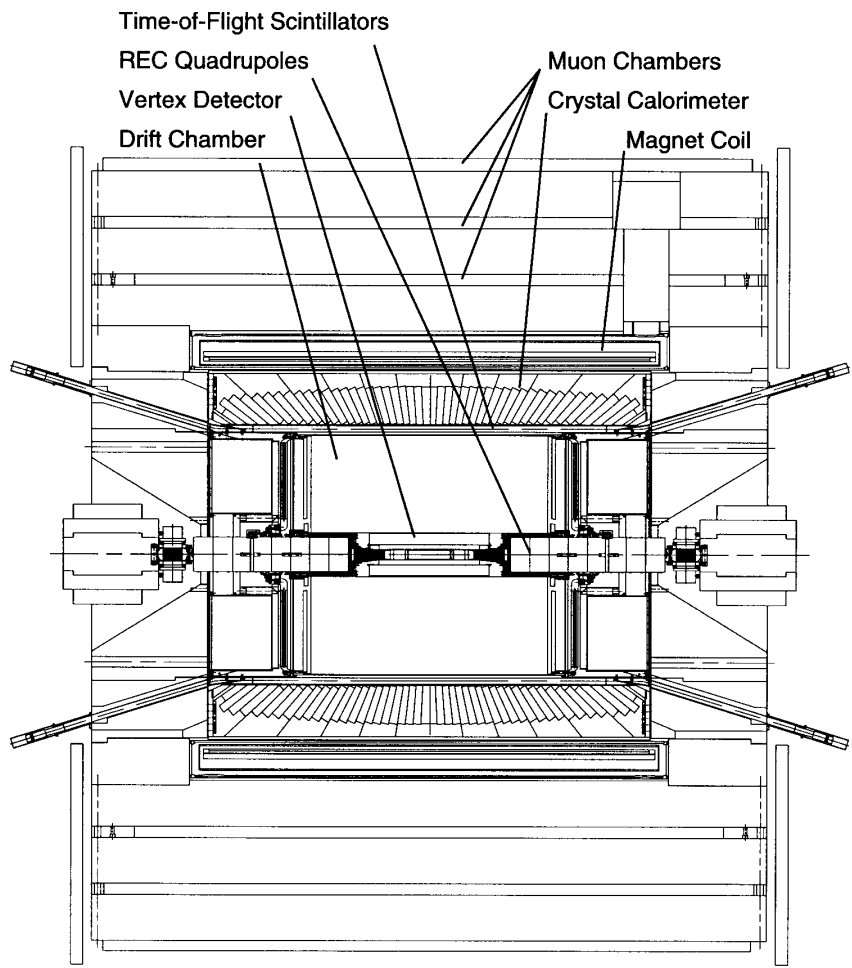,height=5.0in}}
\caption{A schematic of the CLEO detector.
\label{fig:cleodet}}
\end{figure}

\begin{figure}
%\vskip 2.5in
\centerline{\psfig{figure=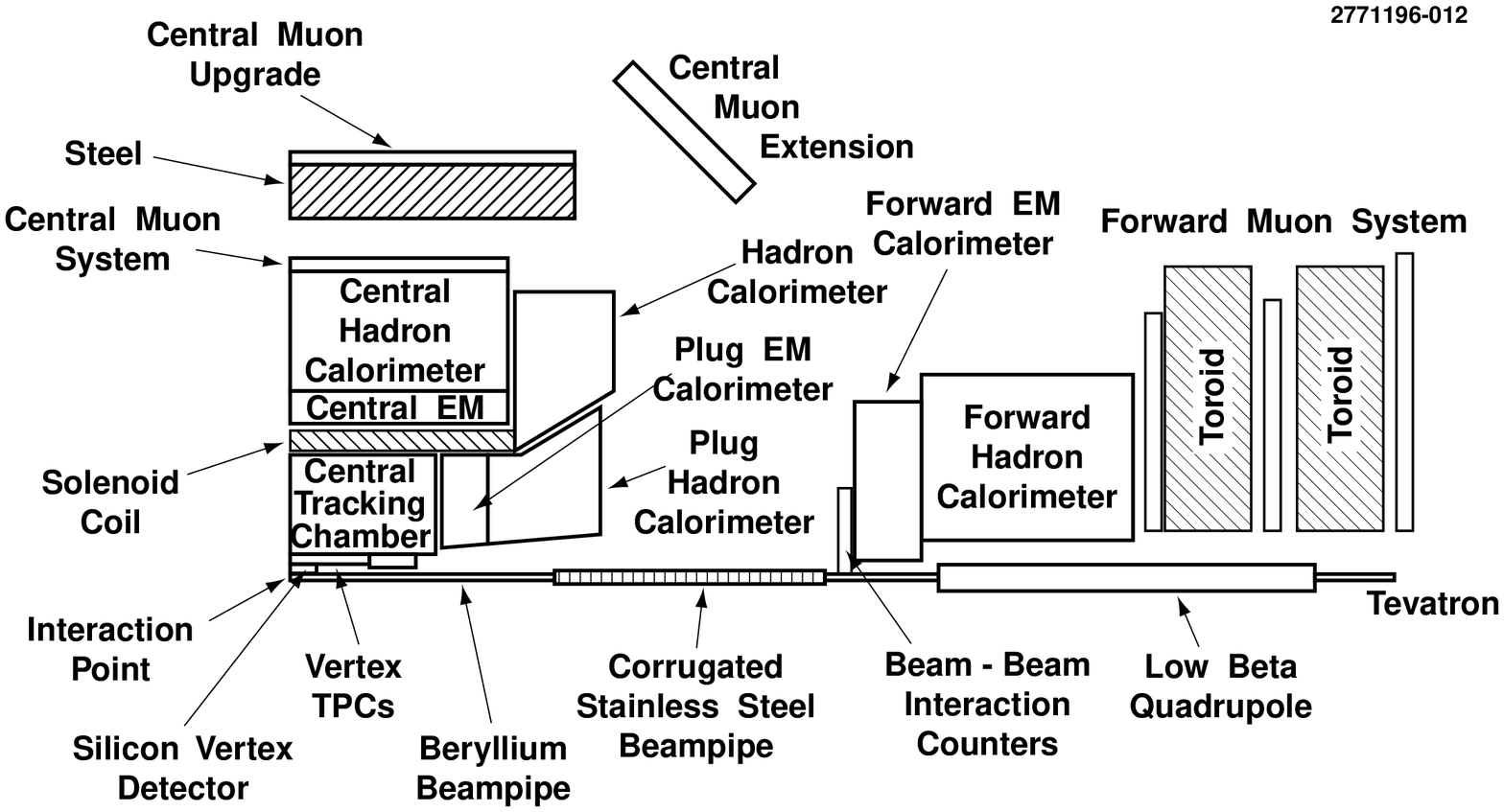,height=2.5in}}
\caption{A schematic quarter view  of the CDF detector showing the
major detector elements. 
(http:\-//\-www\--cdf\-.\-fnal\-.\-gov\-/\-ex\-per\-i\-ment\-/\-drawings\-/\-bw\- quad\-rant \-runi\-.\-ps)
\label{fig:cdfdet}}
\end{figure}

Figure~\ref{fig:alephdet} shows the ALEPH detector which operates at the
LEP $e^+e^-$ storage ring with $E_{CM}= 92$GeV.  
The CLEO detector, which operates at 10.58 GeV in the CM, is shown in
Figure~\ref{fig:cleodet} and CDF, which runs at the 1.8 TeV $p \bar p$ collider
at FNAL, is shown in Figure~\ref{fig:cdfdet}.  As you can see, the three 
detectors are very similar in many ways, but each is individually optimized
to the physics opportunities at its particular machine.

\subsection{Detector Operation}

When the beams at the accelerator collide (or a kaon decays or a $\nu$
interacts in a fixed target experiment), physics, as I said, happens.
The first thing that the experiment has to decide is whether or not an
interaction of interest has occurred at a particular beam crossing or beam spill. This
decision is crucial. If something interesting happens, then the event will be
read out, which takes time (meaning subsequent events will be missed).
In the trade, the process by which the experiment decides whether or not an
event is
interesting is called the trigger. Too loose
or indiscriminate of a trigger will result in lots of
dead time for the experiment so good data will be lost. Too tight 
or selective of a trigger
means interesting physics may be thrown away.

For $e^+e^-$ machines, triggering for most types of events is quite straight
forward. Cross sections are {\it low}. Fairly simple requirements requiring
evidence of a minimum number of charged tracks in the detector or a minimum
threshold for energy deposited in the calorimeter will yield a trigger that is
essentially without deadtime but still preserves 99\% efficiency for $e^+e^-$
annihilation events.

For $p\bar p$ machines and fixed target experiments, the trigger is difficult
and must be carefully thought through. Interactions occur at FNAL almost every
beam crossing. Great care must be taken to suppress unwanted background but
still preserve the $W, Z, $ and $t$ events one wants to study. Kinematics helps
because heavy objects will not have significant boost in the
lab frame. When a heavy object ($Z, W,
t$) is produced, its decay products can have a lot of energy transverse to the
beam direction, while the uninteresting events send most of the beam energy down
the beam pipe.

To give a quantitative comparison, the
total annihilation cross section at LEP is $\sigma(e^+e^-)  \sim 32 $nb. 
At FNAL, $\sigma(p\bar p) \sim 50$ mb (6
orders of magnitude greater).

\begin{figure}
%\vskip 5.5in
\centerline{\psfig{figure=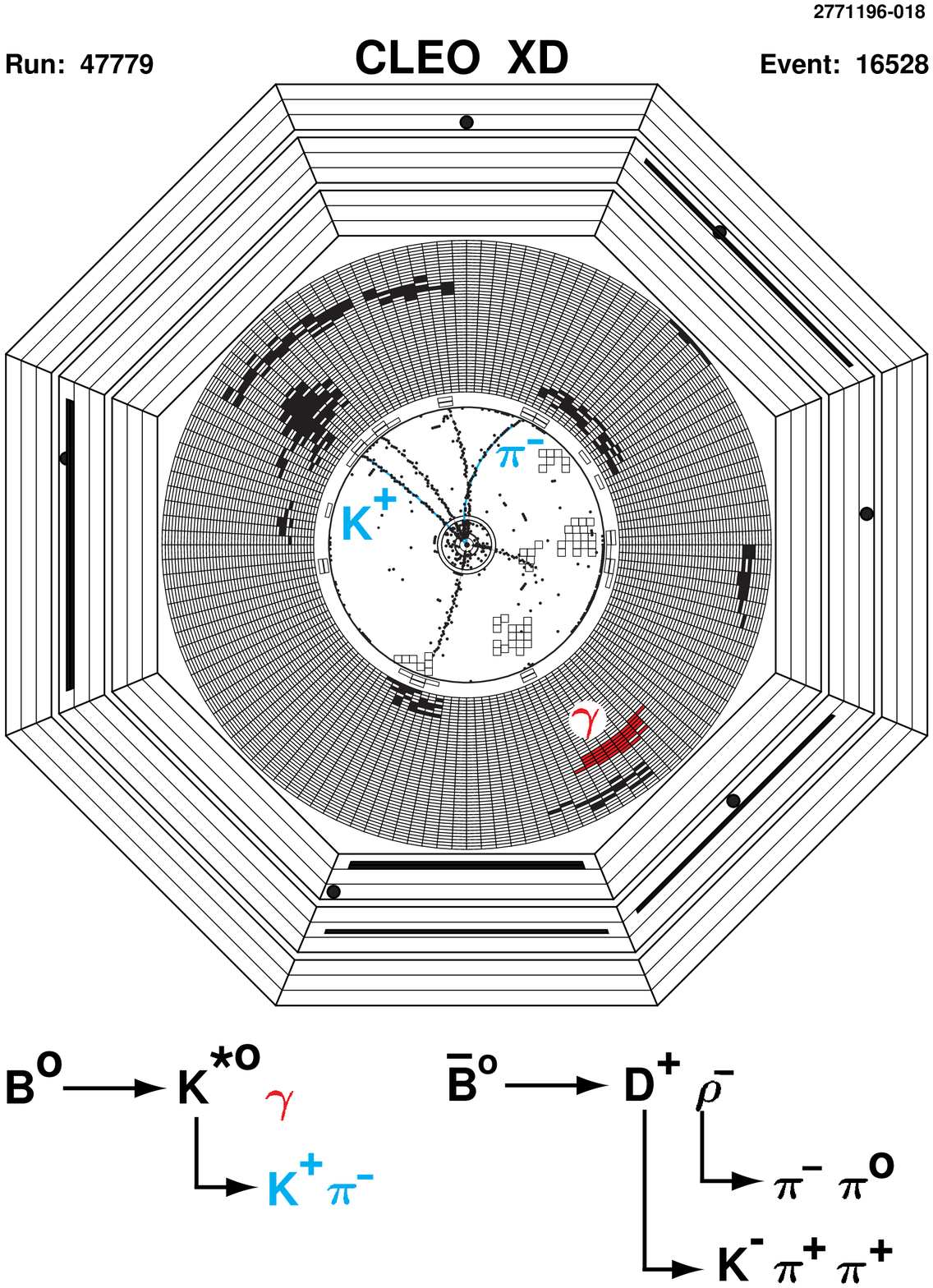,height=5.5in}}
\caption{An event picture from CLEO showing two reconstructed $B$ meson decays.
\label{fig:eventsa}}
\end{figure}

\begin{figure}
%\vskip 3.25in
\centerline{\psfig{figure=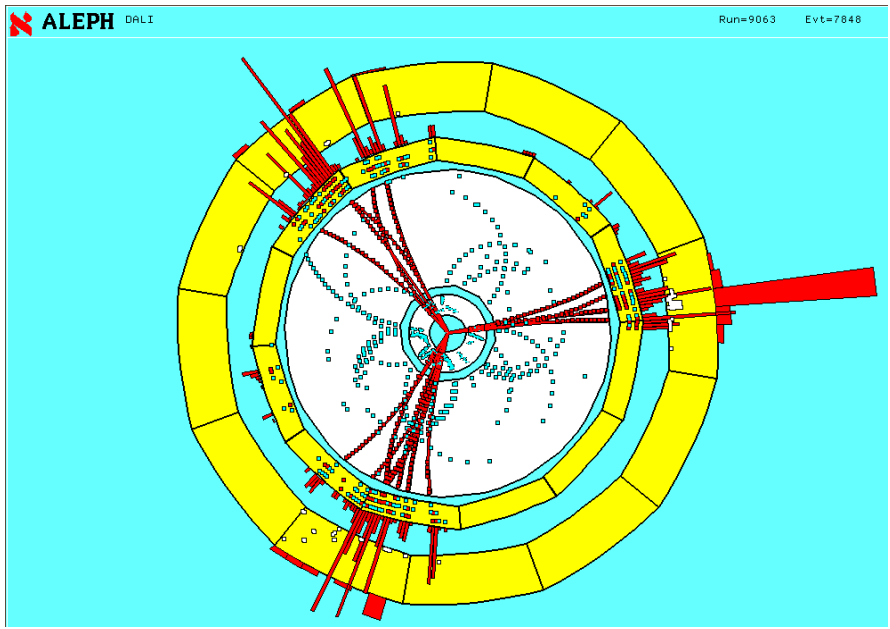,height=3.25in}}
\caption{An event picture from Aleph showing a $Z$ decay to two quarks making two jets and with hard gluon radiation making a third jet. (http:\-//\-alephwww\-.\-cern.\-ch\-/\-WWW\-/\-dali\-gif\-/\-dc009063 \-067848\- cal \-yel \-2 w.\-gif)
\label{fig:eventsb}}
\end{figure}

\begin{figure}
%\vskip 5.5in
\centerline{\psfig{figure=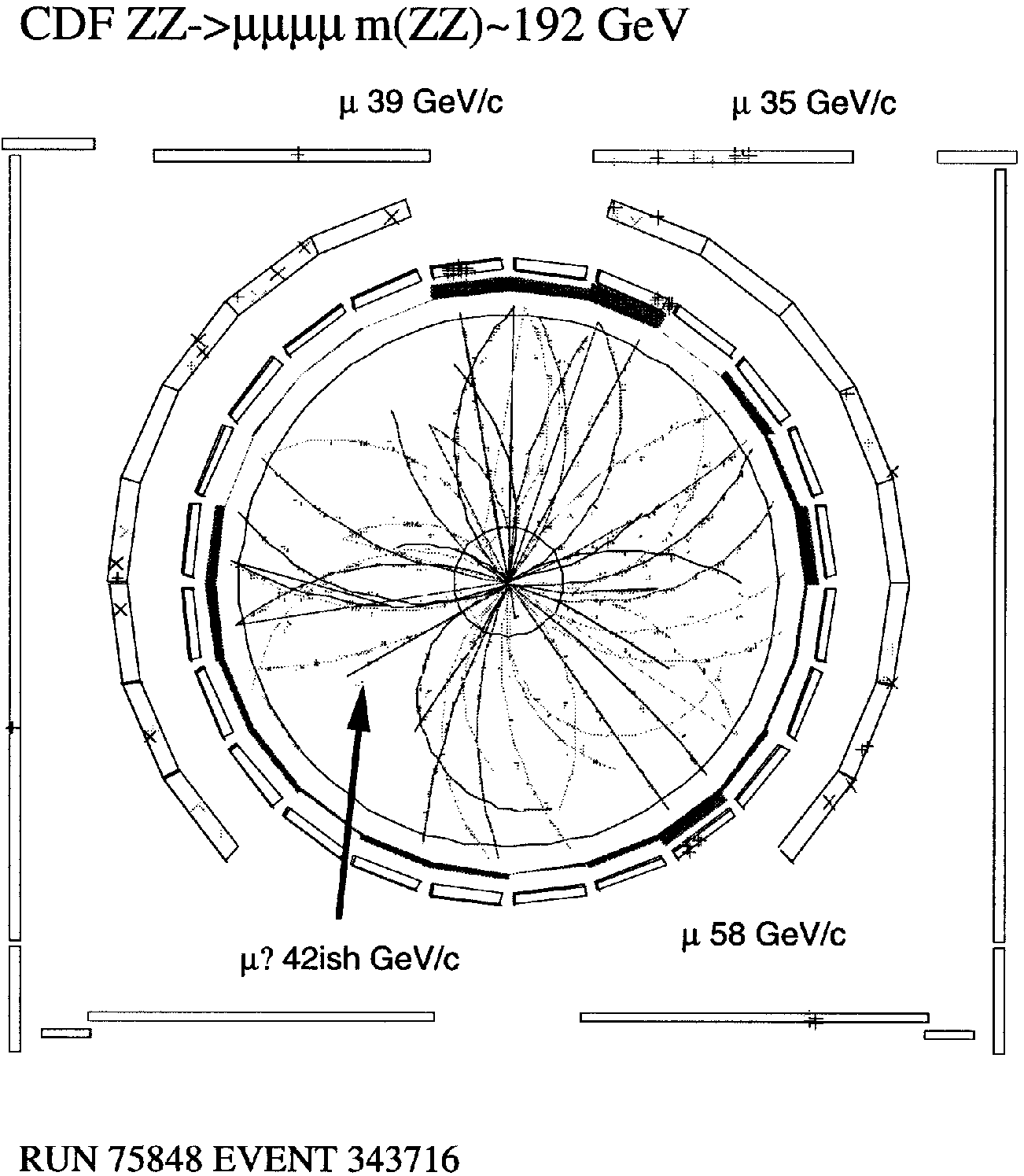,height=5.5in}}
\caption{An event picture from CDF showing a $ZZ\to \mu\mu\mu\mu$ decay. Three of the muons are identified and momenta of the tracks are given.  (http:\-//\-www-\-cdf\-.\-fnal\-.\-gov\-/\-phy\-sics\-/\-ewk\-/\-mmmm.ps)
\label{fig:eventsc}}
\end{figure}

Once the detector is triggered, we read out events.  In Figures~\ref{fig:eventsa},
\ref{fig:eventsb}, and ~\ref{fig:eventsc},
I show a $B$ meson decay from CLEO, a $Z$ decay from ALEPH and a $Z$ decay
from CDF.  The underlying physics process in these events could not have been
identified by looking at the event pictures alone. 
It is only by a careful selection process that
these events can be identified.  However, it is an amusing
exercise to speculate what is going on in individual events.  
Looking at event pictures is fun, instructive, 
and keeps us attached to the real
world, but it is not how we do physics!

\subsection{Data Analysis}
A typical experiment may have  millions of events recorded. A typical physics
analysis may end up with a few hundred events. An analysis searching for a rare
process may end up with a sample of only 10 or 20 events.
One has to develop a procedure to select events characteristic of the physics
process one wants to study but without unnecessary bias. This is an
extraordinary challenge, especially when one considers the magnitude of the
winnowing that must occur.

The primary tool that experimenters have to help them develop a selection
procedure is called ``the Monte Carlo'' (MC).
The Monte Carlo has two parts: the  physics simulation and
the detector simulation.

\subsubsection{Monte Carlo}

Starting from a differential cross section that describes our best understanding
of the physics happening at a given CM energy, an event generator will generate
momentum 4 vectors for a properly distributed sample of events. For example, if we
were interested in studying $Z$ decays at 92 GeV, the physics MC would generate
$Z$ bosons and then decay them to the correct proportions of leptons and quarks
according to whatever model (such as the Standard Model) we specified.

\begin{figure}
%\vskip 3.25in
\centerline{\psfig{figure=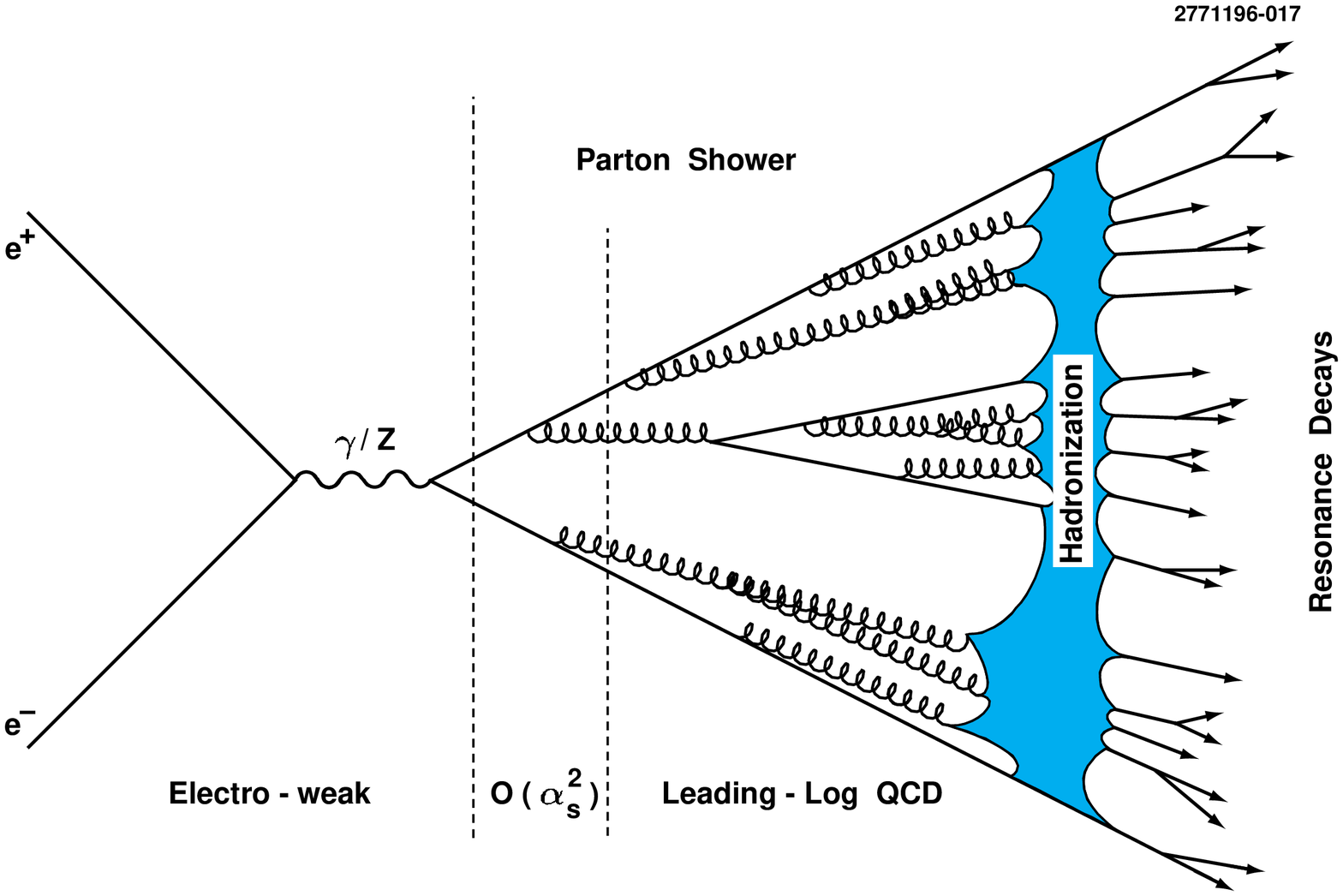,height=3.25in}}
\caption{The fragmentation process is illustrated where partons evolve into
the particles we detect in our detector. (http:\-//\-aleph\-www.\-cern.\-ch/\-ALEPH\-GENERAL\-/re\-ports/\-fig\-ures/\-qcd\-/par\-tsh.gif)
\label{fig:fragment}}
\end{figure}

The Standard Model tells us how to distribute the  4 vectors of quarks and
leptons. We then need some model of hadronization to give us the physical
mesons produced, and then the mesons are decayed according to whatever we
know about their branching ratios and lifetimes.
This procedure keeps going until one has a set of 4
vectors for long-lived particles that will actually end up in the detector,
and is schematically illustrated in Figure~\ref{fig:fragment}.

A word of warning: the physics MC is only as good as the physics we put into it!
If we have neglected some physics in the MC that is present in the data, we will
get discrepancies between what the MC thinks we should be seeing in
our data, and the data we collect. It is always important
to understand what the limitations in the physics inputs to
simulations are.

The second part of the MC is used to simulate how events will appear in the detector
and it is called the detector simulation.
Here one takes the  4 vectors of the stable (long lived)
particles produced by the physics simulation and propagates them through the detector.
The detector simulation, for example, simulates the multiple Coulomb scattering
and energy losses  as the particle passes
through the beam pipe. It propagates the particle through the drift chamber
and generates MC data as simulated signals on drift chamber sense wires.
It simulates the EM shower in the calorimeter and so on.
The detector simulation is hugely expensive in terms of computer time.

The great value of the MC is that one can generate a sample of ``fake data'' or
``MC data'' to test an analysis procedure on. One can determine the effect that
analysis selection
criteria will have on efficiency, one can study potential backgrounds,
and far and away the most important function of MC is that one can, in an
unbiased way, come up with criteria to select a signal.

It is appallingly easy when one is looking for rare processes with small numbers of
signal events and with large backgrounds to end up enhancing a statistical
fluctuation. I will show you some published examples in a few pages. The only
way to avoid that is to use MC data to determine event selection and background
suppression techniques {\it before} ever looking at the data.

\subsubsection{Sample Analysis}

I am going to illustrate for you how an analysis proceeds. I am going to choose
an example of an analysis to measure the rate for a $B$ meson to decay to
the final state $D^{*}\ell \nu$. I choose this particular example because
I will use this decay rate in the
next section as an example of how to measure CKM
mixing angles.
For experimental reasons, we use the decay chain: $D^{*+} \to 
D^0\pi^+, D^0 \to K^-\pi^+$. The experimental quantity that is measured
is the branching ratio
which can be related to the decay rate by the measured $B$ lifetime:
\begin{equation} 
{\rm Br}(B\to D^{*}\ell \nu) =
\frac{\Gamma(B\to D^{*}\ell \nu )}{\Gamma (B\to {\rm all})}
= \tau_B \Gamma(B\to D^{*}\ell \nu)
\end{equation}

\begin{itemize}
\item{\bf Event Selection:}
During this stage of the
analysis, we come up with criteria for selecting specific
events to study.
In the case of $B\to D^*\ell\nu$, we look for events with a $D^*$ and a
lepton 
 in them, with
kinematics consistent with coming from $B\to D^*\ell\nu$ decay.
We use the MC to study both signal events (for which
we want a high efficiency) and background events (which we want to suppress)
and optimize our selection criteria accordingly.
\item{\bf Determination of Backgrounds:}
For this analysis, we may have $D^*\ell$ pairs that are
not from $B\to D^*\ell\nu$ decays.
We can use the MC to help evaluate the
backgrounds, but unless the MC is a perfect description of $B$ decay,
we cannot trust it to absolutely predict the background rates.  Therefore we
try to evaluate as many backgrounds as possible using the data.
\item{\bf Efficiency:}
To evaluate the efficiency of our selection
criteria, we generate $B\to D^*\ell\nu$ events using MC and
pass them through our detector simulation.
We then analyze these MC events the same
way as we analyze data.  It is reasonable
to ask:  why trust the  event generator? 
We don't.  We need to vary the physics generator over
the acceptable parameter space and see how the efficiency of the
analysis is affected. 
Similarly, why trust the
detector simulation? We don't.  We  tune it and test it on data.

\item{\bf Result:}
When we do the analysis, we find a number of events 
N$_{\rm S+B} = 457\pm23\pm9$, where $\pm23$ is the statistical error and $\pm9$
is the systematic error on different ways of extracting the yield.
Of those events, when we subtract backgrounds we find a number of signal
events:
N$_{\rm S} = 376\pm27\pm16$. Note that the statistical error $\pm 27$ has
increased due to the statistical uncertainties in the background subtraction, and
the $\pm16$ systematic error has increased due to modeling uncertainties in the
background.

The final result for the branching fraction is the number of signal events
we observe divided by the number of parent $B$ mesons in our data and divided
by the efficiency of the selection procedure.  We find:
\begin{eqnarray}
{\cal B}(B \to D^*\ell \nu) & = &\frac{N_S}{N_B\epsilon} \\
 & = & [4.49 \pm 0.32 \pm 0.39] \%
\end{eqnarray}
where $N_B$ is the total number of $B$ mesons produced in our data and 
$\epsilon$ is the efficiency for selecting the $N_S$ signal events.

The first error in the result is the statistical error and it depends on
the number of events in the sample and tells the significance of the result.
Most experiments require a result be at least 3 statistical error bars from
a null result before claiming discovery.  The second error is the systematic
error and it is a measure of the stability of the result with changes in the
analysis selection criteria.  Evaluating the systematic error is always the
most difficult and time consuming part of any analysis.  
My personal rule of thumb is that for a result that claims better than 15\%
statistical precision, I am suspicious that the systematic error has not
been properly evaluated if the systematic error is quoted to
be smaller than the statistical error.
\end{itemize}

\subsection{What Can Go Wrong}

\begin{figure}
%\vskip 3.5in
\centerline{\psfig{figure=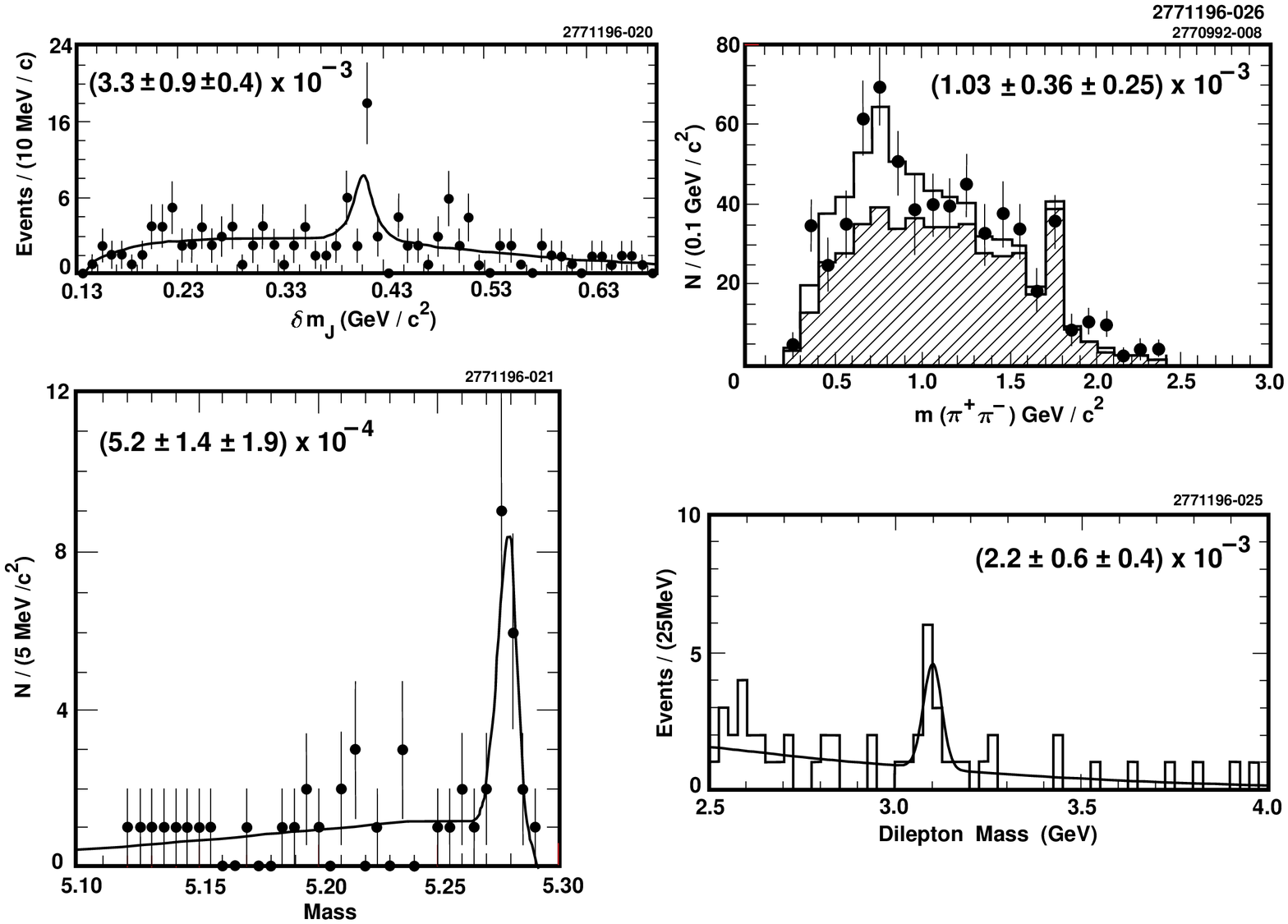,height=3.5in}}
\caption{Plots of three wrong and one correct result.  The results and the
errors are quoted on each plot.  References are deliberately not provided.
The correct plot that is showing a real signal is listed in the text.
\label{fig:wrong}}
\end{figure}

In Figure~\ref{fig:wrong} I show plots of
four experimental results.
Two were
published, one is on its way to being published, and one was retracted before being
published. Three of the four are wrong and the signal that they are 
supposedly demonstrating evidence for does not exist at a level consistent
with the claims of the analysis. Can you tell which are wrong and which is right?
(To protect the guilty, I am deliberately not going to provide references for
the four plots.  Each was a measurement of a branching ratio, and I have 
quoted the numerical value measured on each plot so that the reader can
see the central value and the errors.)

The point of showing this
figure is that it is not at all obvious
by looking at the plots which is right or wrong.  One needs to examine
the individual analyses in more detail.  It is important to ask:
What are the pitfalls? Where do experimenters make mistakes? How can you tell?

In two of the flawed results of Figure~\ref{fig:wrong}, my
personal opinion is that the selection cuts for a signal were tuned on the
data instead of on MC. In a third result, the mistake was, I believe,
 a large background
that the experimenters assumed the MC modeled properly and it didn't.

The correct result is the top left plot of Figure~\ref{fig:wrong} and
I have deliberately shown the worst looking plot from the analysis. The result
can be made to look much
better with different binning. That is considered cheating if it
affects the signal yield extracted by the analysis. In this
particular analysis, the yields were
computed with very fine binning and it was
tested that they were independent of bin size. The evaluation of the
background
was done many different ways (from data and MC ) and the result was
stable when the cuts were changed. These are all checks you should expect to see
experimenters do.

So what are the questions you should ask when deciding whether to believe a
marginal (3--4 $\sigma$) result or in deciding whether you believe the level of
precision on a more significant result?
\begin{enumerate}
\item How were event selection criteria determined?
\item How was the background evaluated?
\item What happens when the event selection cuts are varied?
\item What is the error on the efficiency and how was it determined?
\end{enumerate}

\noindent
There are some other pitfalls you should be aware of:
\begin{itemize}
\item Sometimes experimenters do not understand their detectors as well as they
think they do.
This is evident if one looks at the high resolution
measurments of the $B$ meson lifetime as a function
of time~\cite{pdg} starting in 1986
as shown in Figure~\ref{fig:blife}.  The
$B$ lifetime has increased by almost 50\% of its value as increasingly
precise detectors have been used and larger data sets have been analyzed.
Some of the early results were optimistic about their error bars and the systematic
shift has presumably resulted from improvements in the analysis procedures and
a better understanding of the detectors.  
\begin{figure}
%\vskip 3.0in
\centerline{\psfig{figure=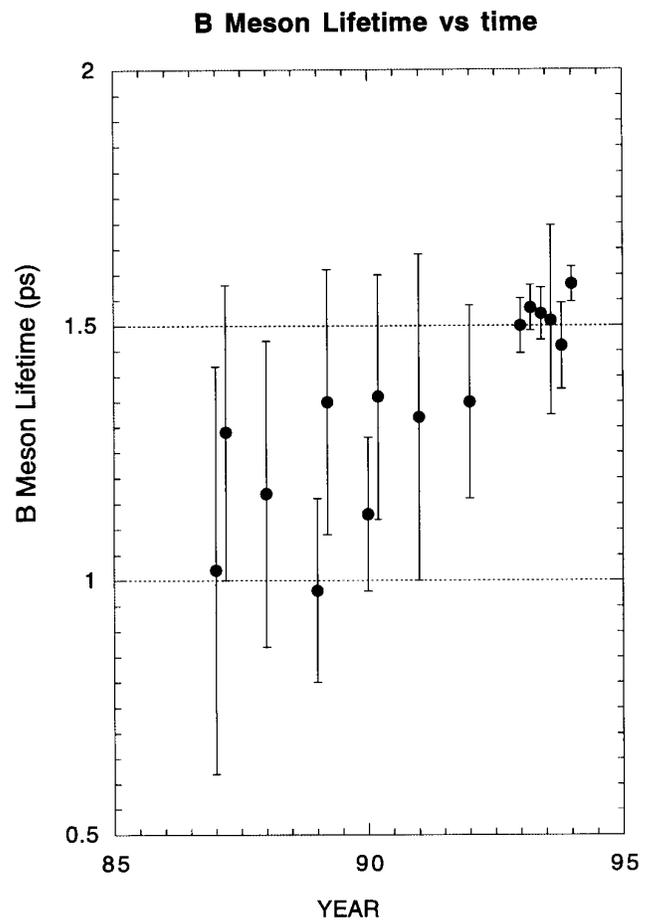,height=4.75in}}
\caption{The measured $B$ meson lifetime as a function of time.
\label{fig:blife}}
\end{figure}

\item Experimentalists and theorists alike are prone
to over-averaging, where many measurements are averaged, weighted by 
their combined
statistical and systematic error. The Particle Data
Group, a wonderful institution in every
respect, contributes to this problem in some ways by making the data so
easily available.

A famous example of over averaging was the ``$\tau$ 1-prong problem.'' The
$\tau$ lepton can decay just like the muon ($\tau\to e\nu\bar\nu$,
$\tau\to\mu\nu\bar\nu$), but it can also decay to hadrons.
In 1984, it was noticed that if one took the world averages for the measured
$\tau$ branching ratios and used theoretical predictions to constrain
 poorly measured modes,
then the measured inclusive 1-prong branching ratio
 was significantly larger than the sum of
the exclusive modes.~\cite{tau} As late as 1992, one saw that, taking world 
averages, one found
 a significant discrepancy in the inclusive rate and
the sum of the exclusive rates.~\cite{dallas}

For years there were speculations about new physics and unseen decay modes.
The  problem, however, was caused by averaging many experiments with large
errors and extracting an average with rather small errors.  It is
very dangerous to take
results from 10 different measurements
with roughly equal precision  and then
average them to get a factor of three
smaller error.
If errors were statistical only, there would not be a difficulty. The
problem comes from systematic errors
 which may be correlated experiment to experiment.
Systematic errors are hard  to evaluate to begin with, 
and correlations are hard to spot.  For example, there may be unknown
correlated errors due to incorrect
inputs to the MC or overlooked backgrounds.
The moral is that global averages need to be done with great care and even then,
I believe that one needs to use a higher 
threshold for claiming a significant discrepancy when averaged data is
being used.

\item A third pitfall that affects experimenters more than theorists, but you
should be aware of it, is the enormous temptation to stop at the
``right answer''.  Certainly anyone who has ever done a freshman physics
lab knows that feeling.  We very often have a
preconceived prejudice on what a result should be.
A good experimenter does many of the systematic studies and
checks {\it before} looking at the actual
number he or she is getting. 

\item Finally, theorists tend to fall into the ``single
event'' pit. What does it mean to find a single event? 
In 1964, the $\Omega^-$ was discovered with the observation of
 one event.~\cite{omega} However, there
was an enormous amount of information in the bubble chamber
photograph that captured that one event.  The decay was fully
reconstructed $K^- p\to\Omega^- K^+ K^0, \Omega^-\to\Xi^0 \pi^-,
\Xi^0 \to\Lambda^0 \pi^0, \Lambda^0\to\pi^- p, \pi^0\to\gamma_1 \gamma_2, 
\gamma_1\to
e^+ e^-, \gamma_2\to e^+ e^-,$ except for the $K^0$.
They were able to claim discovery because there
was so much information in the event that the  probability for the
background to produce such an event was vanishingly small.
However, for modern collider experiments, it is
impossible to have the same level of information,
especially in $p\bar p$ collisions where much of the event goes down the beam pipe.
The crucial issue is not how many events one
finds, but how well the background can be evaluated and understood, and
what is the probability that a background process could imitate the
event one is looking for.
\end{itemize}

\section{Measuring Parameters of the Standard Model}

In Section 2, I listed the parameters of the Standard Model that must be
determined experimentally. I will spend this section discussing some of those
experiments. I like to think of them as the measurements that define the Standard
Model.

\subsection{Measurements of Coupling Constants}

I will start with a description of how to measure a coupling constant.
The coupling constants in the SM are $\alpha_{\rm EM}, \alpha_s$ and $G_F$.
 The most precise determination of $\alpha_{\rm EM}$
comes from the electron $g-2$ experiments 
using single trapped electrons~\cite{pdg}:
\begin{equation}
\alpha_{EM}^{-1} = 137.035\ 992\ 35\ \left(73\right)
\end{equation}
$G_F$ is determined from the muon mass and lifetime using the relation:
\begin{equation}
\frac{1}{\tau_\mu} = \frac{G_F^2M_\mu^5}{192\pi^3}
\end{equation}
and the dominant uncertainty on $G_F$ comes
from the second order radiative corrections to this expression~\cite{pdg}, and the
result is:
\begin{equation}
\frac{G_F}{\left(hc\right)^3} = 1.16639\left(2\right)\times 10^{-5} {\rm
GeV}^{-2}
\end{equation}
$\alpha_s$ is the most poorly measured quantity in the entire physical
constants list of the Particle Data Group~\cite{pdg}:
\begin{equation}
\alpha_s(M_Z) = 0.116(5)
\end{equation}
There is some theoretical uncertainty over how best to determine $\alpha_s$; I expect
lots of progress here in the next few years.

\begin{figure}
%\vskip 3.15in
\centerline{\psfig{figure=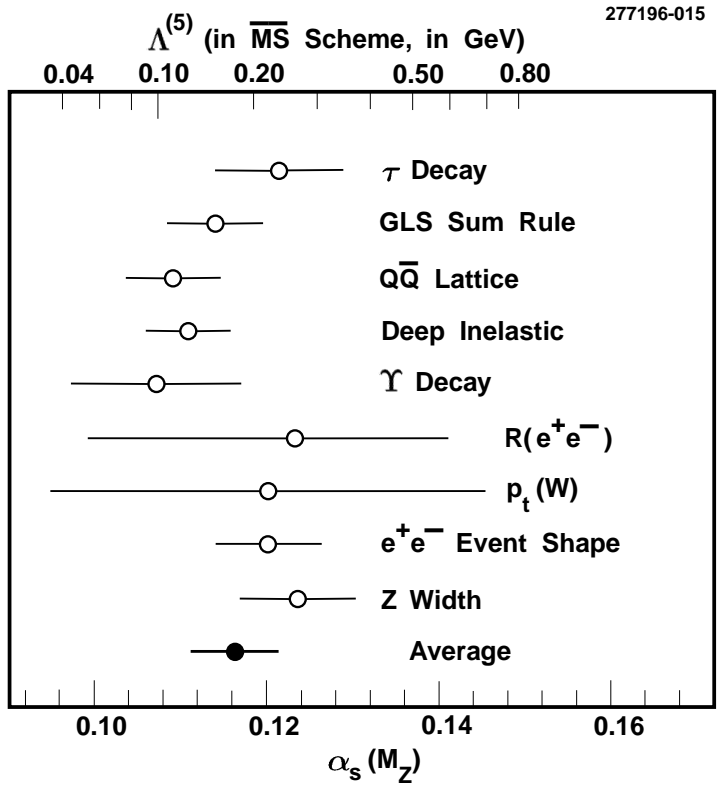,height=4.15in}}
\caption{A plot of the value of $\alpha_s$ as measured in a variety of
experiments at different $Q^2$.  All measurements have been evolved to
$\alpha_s(M_Z)$ for comparison purposes.
\label{fig:alphas}}
\end{figure}

I want to talk briefly about how $\alpha_s$ is measured. There are
many different
techniques that are employed at different $Q^2$ 
as shown in Figure~\ref{fig:alphas}, and the
remarkable agreement between them is considered one of the great successes of QCD~\cite{pdg}.

The most obvious way to measure a coupling constant of a particular interaction is to
measure the energy spectrum of the system bound by that interaction. For
example, in the early days of quantum mechanics, 
the Rydberg was measured from the spectrum of atomic
hydrogen. Similarly, one uses the spectroscopy of a system bound by the 
strong interaction to measure
$\alpha_s$.
This procedure has a slight difficulty. For QED, the interaction that binds the
$e^-$ in the hydrogen atom, we can write down the Schrodinger equation and solve it
 to get the relation between the measured energy levels and the coupling constant of the
interaction.  For QCD it is not so easy; there is no equivalent to the Schrodinger
equation for the strong interaction.  However, QCD is being solved with lattice
techniques, allowing us to relate $\alpha_s$ to the measured energy splittings~\cite{lepage}.

\begin{figure}
%\vskip 3.0in
\centerline{\psfig{figure=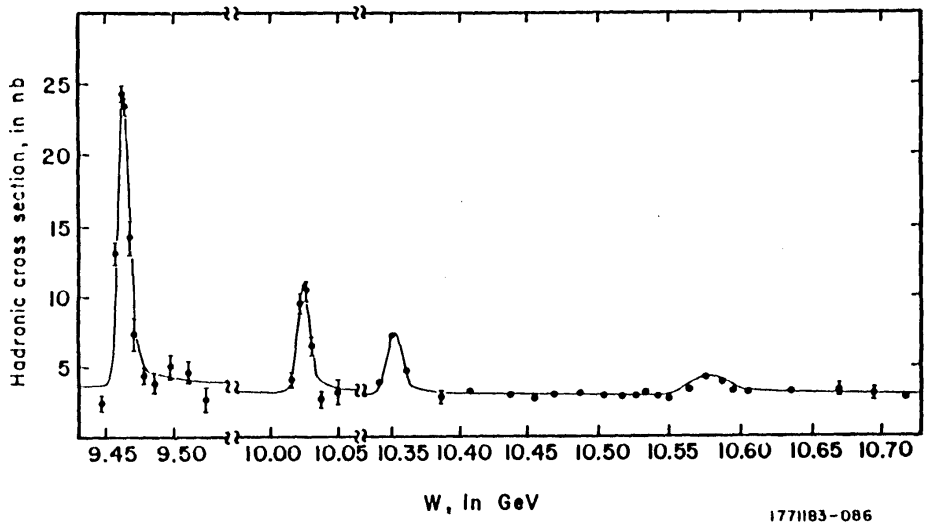,height=2.75in}}
\caption{A scan of the hadronic cross section versus energy in the $\Upsilon$
energy region from CLEO showing the $b \bar b$ bound states.
\label{fig:upsilon}}
\end{figure}

One of the bound state systems used for the $\alpha_s$ extraction is the $\Upsilon$
system (a bound state of a $b$ and a $\bar b$ quark).
The mass of the $b$ quark is around 5 GeV. Between the energies of 9.46
and 11 GeV, the spectrum of $b\bar b$ quark bound states is rich.
In an $e^+ e^-$ machine such as CESR, the bound states with the same quantum numbers
 as the
photon are copiously produced and show up as dramatic features in a scan of
hadronic cross section versus CM energy as shown in Figure~\ref{fig:upsilon}.

The experiment is easy to do. One scans the energy of the $e^+ e^-$ beams and looks at the
number of events in the detector, where an
event is  loosely defined as three or more tracks.

There is an underlying continuum of $e^+ e^- \to \gamma^* \to q\bar q$ events. 
Then there is a dramatic
increase in the number of events observed when producing the $^3S_1$ states of 
the $b\bar
b$ bound state system.
The observed resonances are the $n=1, n=2, n=3$, and $ n=4$ radial excitations.
The low-lying
resonances are narrow because their decays,
which are dominantly three gluon exchange,  are suppressed. The  $^3S_1$, $n=4$ radial
excitation is wide because it is massive enough to be fully allowed to strongly decay
to a pair of $B$ mesons.

From the energy of the $e^+ e^-$ beams, we get the
energies of  the $\Upsilon$ states.
Other bound states of the $b\bar b$ system are observed by seeing pion or photon
transitions from the $^3S_1$ states.  Similar studies are performed 
on the $c\bar c$ bound states such as the $J/\Psi$. By fitting the energy levels,
one gets~\cite{pdg}
\begin{equation}
\alpha_s\left(M_Z\right) = 0.110 \left(6\right)
\end{equation}
where it is convention to evolve $\alpha_s$ from the $Q^2$
where it is measured up to $M_Z$ using the renormalization group equations.
The errors are dominated by the systematics
 of the lattice calculation: a finite lattice spacing is used and
the quenched approximation is made, where no light quarks are allowed to propagate.

Other ways exist to measure $\alpha_s$. In almost all other methods, the
measurement is sensitive to $\alpha_s$ as a radiative correction. As an example,
\begin{equation}
R = \frac{\sigma\left(e^+e^- \to {\rm hadrons}\right)}{\sigma\left( e^+e^- \to
\mu^+\mu^-\right)} \end{equation}
can be used to measure $\alpha_s$.
You have all calculated the cross section for $e^+e^- \to \mu^+\mu^-$
\begin{equation}
\sigma_{\mu^+\mu^-} = \frac{4\pi\alpha_{EM}^2}{3s}, \qquad s=E_{CM}^2
\end{equation}
For $e^+e^- \to q\bar q$ the $\mu$ charge ($e$) is replaced by $Q|e|$ and you
get an extra factor of 3 for 3 quark colors.
\begin{equation}
\sigma_{q\bar q} = \frac{4\pi\alpha_{EM}^2}{3s}Q_i^2\times 3
\end{equation}
Naively, then, $R = 3\cdot\sum Q_i^2$ since the quarks will make hadrons
100\% of the time.
However, the expression for $R$ is modified by higher order QCD corrections to be
\begin{equation}
R = R^{\left(0\right)}\left[1+\frac{\alpha_s}{\pi}+....\right]
\end{equation}
and so by careful experimental measurements of $R$ one can extract $\alpha_s$.

At the moment theoretical errors dominate in virtually all measurements of
$\alpha_s$.

\subsection{Measurements of a Gauge Boson Mass}

Once $G_F$ and $ \alpha_{EM}$ are
 measured, we need one more experimental quantity to define the SM,
and then all other measurements of fermion couplings, $W$ mass, and so on, will
constitute checks of the model.
We need to determine  either a gauge boson mass ($M_Z, M_W$) or a weak
mixing angle ($\sin^2\theta_W$) from deep inelastic scattering or atomic physics.
We want to use the
 most precise quantities available to define the  model. Ever since
the precision  LEP measurements of the $Z$ mass, $M_Z$ has become the
third parameter of choice.

In principle and in practice, you can measure $M_Z$ in either $e^+e^-$ or $p\bar
p$ collisions. In $p\bar p$ collisions, 
there is a broad spectrum of incoming parton energies and the initial
state energy is not known.  However, one can
use 
the clean leptonic decays $Z \to e^+e^-, \mu^+\mu^-$ to select background-free
samples of
events, and from a measurement of the momentum vectors of the final state particles 
(typically with few percent
resolutions), one can reconstruct the invariant
mass of the parent boson and determine $M_Z$.

In $e^+e^-$ collisions, the CM annihilation energy is
well known. The machine energy spread (the energy spread of the
electron and positron beams) is much less than the width of the resonance.
One can study the
resonance shape directly.  There is very little background and these experiments
have the highest precision.  

The cross section to produce $Z$'s at the pole is
large (30 nb). A machine like LEP can produce
thousands of $Z$'s per day and the very large
data samples have made detailed studies of all the decays of the $Z$ possible.
It is again straightforward to calculate
 $\sigma\left(e^+e^- \to Z^0 \to f\bar f\right)$ in the SM~\cite{quigg}.
 At the $Z$ pole, the contribution to  $\sigma\left(e^+e^- \to f\bar
f\right)$ from QED is negligible.
\begin{equation}
\sigma_{\rm pole} \left( e^+e^- \to Z^0 \to f\bar f \right) = \frac{G_F^2M_Z^4
s\left(R_e^2+L_e^2\right)}{24\pi\left[\left(s-M_Z^2\right)^2 +
M_Z^2\Gamma_Z^2\right]} \left(R_f^2 + L_f^2\right)
\end{equation}
Here, $R_e$ and $L_e$ are the right and left handed electron couplings to the $Z$, 
$R_f$ and $L_f$ are the right and left handed fermion couplings to the $Z$, and $\Gamma_Z$
is the total width of the resonance.  This is
often expressed using
\begin{equation}
\sigma_0^{f\bar f} = \sigma\left(\sqrt{s}=M_Z\right) =
\frac{G_F^2M_Z^4}{24\pi\Gamma_Z^2}\left(R_e^2 + L_e^2\right)
\left(R_f^2 + L_f^2\right)
\end{equation}
which then gives:
\begin{equation}
\sigma_{\rm pole} \left( e^+e^- \to Z^0 \to f\bar f \right) = \sigma_0^{f\bar
f} \frac{s\Gamma_Z^2}{\left[\left(s-M_Z^2\right)^2 + M_Z^2\Gamma_Z^2\right]}
\end{equation}

$M_Z$ determines the location of the Breit-Wigner
resonance, $\Gamma_Z$ determines the width, and $\sigma_0$ determines
the normalization.
In fact, only one free parameter 
is needed to fit the $Z$ line shape: $M_Z$.  The SM
predicts the values of $\Gamma_Z$ and $\sigma_0$ in terms of $M_Z$.
Initially, however,
one fit the resonance to three independent
parameters, $\sigma_0, \Gamma_Z,$ and $ M_Z$ to check the model for consistency.

It is impossible to go further in
our discussion of $M_Z$ without a discussion of
radiative corrections.
In the study of the $Z$ resonance, there are
two types of radiative corrections.
\begin{itemize} \item QED radiative corrections:  here real photon emission from
an
initial state $e^+$ or $e^-$ occurs. 
It is a dramatic effect, and does not contain
any particularly new or interesting physics.
\item EW radiative corrections:  these come
in as vacuum polarization and vertex corrections to the
tree level process $e^+e^- \to Z^0 \to f \bar f$. These
corrections affect the tree level relations
between $\alpha_{EM}, G_F, M_Z$ and $M_W, \sin^2\theta_W$ derived from other
experiments. 
\end{itemize}

\subsubsection{QED Radiative Corrections}

When you tune $e^+e^-$ beams to a particular CM energy, you want to measure the
cross section at that energy. 
In fact, however, you
may be sampling the cross section at some other lower energy because
bremsstrahlung from  the
incoming $e^+$ or $e^-$ has removed energy from the CM. In an
experiment, one actually samples the  entire
cross section below the nominal CM energy with a sampling spectrum
$f\left(k\right)$ that is determined by the physics of bremsstrahlung.
\begin{equation}
\sigma_{\rm obs}\left( E\right) = \int dk f\left(k\right) \sigma_{\rm
BW}\left[ E\left( E-k \right)\right]
\end{equation}

There are two reasons that this becomes a large effect at the $Z$ resonance.
\begin{enumerate}
\item The amplitude for single photon emission 
from the electron or positron can be written in terms of the annihilation
amplitude with no photon emission as:
\begin{equation} 
a_{1\gamma} = e \left( \frac{p^+\cdot \epsilon}{k\cdot p^+}
- \frac{p^-\cdot \epsilon}{k\cdot p^-}\right) a_{0\gamma}
\end{equation}
where $k$ is the photon 4-vector, $p^\pm$ refers to the electron and positron
4-vector, and $\epsilon$ is the photon polarization.  
Summing over photon polarization and integrating over photon angles we get
the
change in the cross section due to initial state radiation:
\begin{equation}
d\sigma = \sigma_0\left(s\right) \left[
\frac{2\alpha_{EM}}{\pi}\left(\ln\frac{s}{m_e^2}-1\right)\right]
\frac{dk}{k}
\end{equation}
where the term in the square brackets is often call $\beta$, the
``effective coupling'' for bremsstrahlung. $\beta$ is
large ($\ln\frac{s}{m_e^2} =
24.2$ at the $Z$ resonance and $\beta = 10\%$) due to the large phase space
 for an $e^-$ to shake off a
nearly collinear photon.

\item To get the cross section, we
integrate $d\sigma$ (to do this right we need to include
vertex corrections to remove infrared divergences as $k\to 0$.)
\begin{equation}
\int_0^{k_{\rm max}} d\sigma = \sigma_0\left( s\right)\left[ 1+\delta_1 +
\beta\ln\frac{k_{\rm max}}{E}\right]
\end{equation}
where $\delta_1$ is often called the first order $e^-$ form factor.

To evaluate this
expression, we need $k_{\rm max}$.  In general bremsstrahlung energies can
extend up to the kinematic limit which is the
 beam energy, but when  one is sitting on a narrow
resonance, if the initial state
radiation is  much more than the  width of the resonance, one is moved
into a region of very low cross section. Therefore, the
resonance cuts off contributions from
hard photons which in effect puts an upper limit $k_{\rm max} \sim \Gamma/2$.
\begin{equation}
\sigma\left(M_Z^2\right) = \sigma_0\left(M_Z^2\right)\left( 1+ \delta_1 + \beta\ln
\frac{\Gamma}{M}\right)
\end{equation}
The narrow resonance cuts off contributions from all but the
softest radiative events, depressing the cross section significantly ($\beta\ln
\frac{\Gamma}{M} \sim -39\%$ at the $Z$), 
and the resonance shape is skewed by a high energy tail.
\end{enumerate}

The
stunning fact is that
it was  only in 1987  that the
second order calculations of these QED radiative
corrections were completed;
just in time for LEP and SLC.~\cite{radcor}

\subsubsection{EW Radiative Corrections}

At tree level in the Standard Model, we measure $\alpha_{EM}, G_F, M_Z$
and from them we can calculate
$M_W$ and $ \sin^2\theta_W$.  We can then check the Standard Model by measuring $M_W$
or $\sin^2\theta_W$ directly.
Unfortunately, life is not lived at tree level, and when we measure the $Z$ mass, 
we really
measure the sum of the tree level process
and all the radiative corrections to it. Nature has summed the
perturbation series for us.
One effect of these higher order corrections is that coupling constants run  and their values change with $Q^2$.
For example, in QED:
\begin{equation} \alpha_{\rm EM}\left( q^2\right)
= \frac{\alpha_{EM}}{1-\frac{\alpha_{EM}}{3\pi}\ln\frac{q^2}{m^2}}
\end{equation}
One way to think of this renormalization of the electron charge is that it comes
from the static polarizability of the vacuum. At higher values of $q^2$, one is probing
shorter distances, getting closer to the bare charge which is infinite.
As experimentalists, we are lucky that nature (correctly)  
has computed all the radiative corrections
for us to all orders.
The problem is that the radiative corrections will modify the
simple tree level relations between,
for example, $G_F$ and $\sin^2\theta_W$ or $\sin^2\theta_W$ and $M_W$.

Consider the
following two definitions of $\sin^2\theta_W$.  They are equivalent at tree level:
\begin{eqnarray} 
\sin^2\theta_W & = &1 - \frac{M_W^2}{M_Z^2} \\
\sin^2\theta_W&  = & \sqrt{\frac{4\pi\alpha_{EM}}{\sqrt{2}G_F M_Z^2}} \label{eq:sinsq}
\end{eqnarray}
where in both expressions, the physical boson masses are used, and
in the second expression, $\alpha_{EM}$ and $G_F$ are determined from
low energy experiments.
These two relations, while equivalent
at tree level, will  give different results when experimental data are used.

We can parameterize the effect of the
radiative corrections by a correction to our relations. We can
define $\sin^2\theta_W$ from the physical boson masses as:
\begin{equation}
\sin^2\theta_W \equiv 1 - \frac{M_W^2}{M_Z^2}
\end{equation}
Then equation ~\ref{eq:sinsq} is modified: 
\begin{equation}
\sin^2\theta_W = \sqrt{\frac{4\pi\alpha_{EM}}{\sqrt{2}G_F M_Z^2\left(1-\delta
r\right)}}
\end{equation}
where $\delta r$ incorporates the effects of the radiative corrections.  The
largest contribution to $\delta r$ is just from the running
of $\alpha_{EM}$ to the  $Z$ mass.
\begin{equation}
\alpha_{EM}\left( M_Z^2 \right) = \left[128.8\pm 0.12\right]^{-1}
\end{equation}
and the next largest contribution to $\delta r$ comes from the top quark 
which gives a contribution:
\begin{equation}
\delta r_t = -\frac{3G_F m_t^2}{8\sqrt{2}\pi^2} \frac{1}{\tan^2\theta_W}
\simeq -0.0102 \left[ \frac{m_t}{100{\rm GeV}/{\rm c}^2}\right]^2
\end{equation}
which gives a 3\% correction for $m_t \simeq 175$ GeV/c$^2$.

The fact that the radiative corrections to the electroweak observables are
sensitive to the top quark mass means that precision measurements of the radiative
corrections can be used to determine the top mass.
If one assumes that the SM is correct and there is
no new physics, then a 
combination of a measurement of $\alpha_{EM}, G_F$, and $M_Z$ with one other
precise measurement (i.e., $\sin^2\theta_W$ extracted from a measurement
of quark or lepton couplings to the $Z$) gives
a  ``measurement''
of $m_t$! Before the top quark
was discovered, LEP  provided a tight ($\pm 15\%$) constraint on its mass
in just this fashion~\cite{lang}.

\subsubsection{Experimental measurement of the $Z$ mass}

\begin{figure}
%\vskip 3.05in
\centerline{\psfig{figure=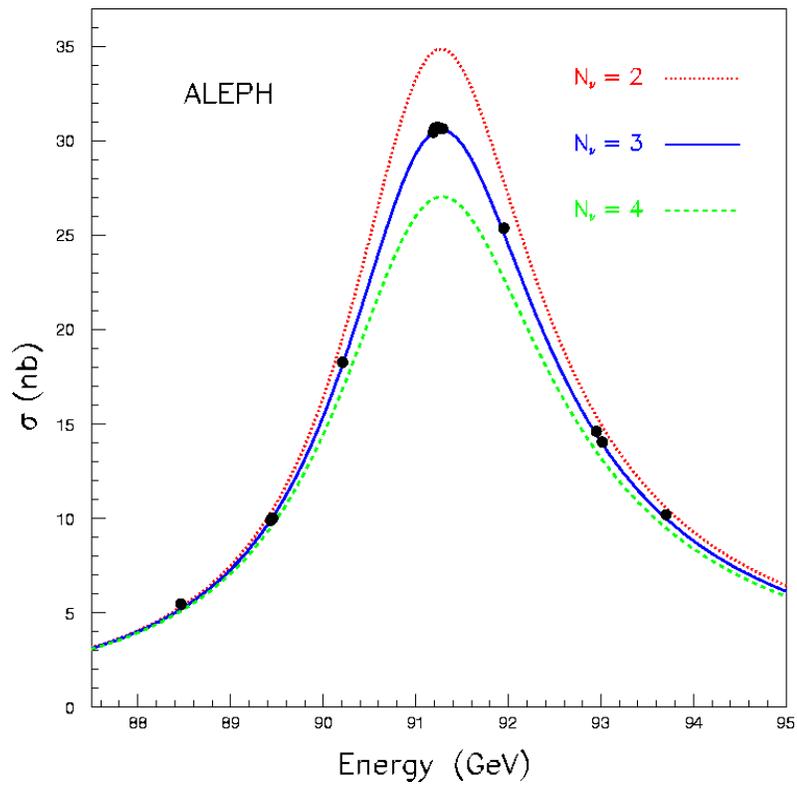,height=4.05in}}
\caption{A scan of the $Z$ resonance taken by ALEPH. The three curves indicate the expected lineshape for 2, 3 and 4 flavors of light neutrinos.  (http:\-//\-www.\-cern.\-ch/\-ALEPH\-GENERAL/\-re\-ports\-/fig\-ures\-/ew\-/z\-line \-aleph.\-gif)   
\label{fig:z}}
\end{figure}

The precision measurement of the $Z$ mass comes from LEP.  
LEP is an electron positron
collider, 26.66 km in circumference, capable, ultimately,  of almost
100 GeV/beam. 
To measure the $Z$ mass, one
scans the energy of the machine through the $Z$ resonance. The
number of events of the types  $Z^0 \to$
hadrons, $e^+e^-, \mu^+\mu^-,\tau^+\tau^-$ in the detector are
counted. A plot of the number of events
versus energy, such as is shown in Figure~\ref{fig:z}, is fit to extract $M_Z$.

The precision of the measurement depends crucially on the absolute
 energy calibration of the machine.
LEP uses a resonant spin depolarization technique with
an intrinsic accuracy of $\sim 1$ MeV to determine the absolute energy
scale of the machine during special experimental runs.  This calibration then
must be transferred from 2 GeV above the $Z$ resonance where the calibration experiment is
done to the energies of the scan points for the mass measurement.
In practice, LEP
achieves a systematic
error of $\pm 3.7$ MeV for the absolute beam energy. In attempting to
improve the systematic error, studies have shown that the energy of the LEP
machine is sensitive to the tides, the lake levels, the magnet temperature, and,
most recently, they have found a correlation between the voltage on the TGV and
the energy of the LEP beam at the MeV level!  For the $Z$ mass measurement, LEP
quotes an
uncertainty on the absolute 
beam energy calibration~\cite{deltae} of $\sigma_E/E = 5.3 \times 10^{-5}$, and
from the scan to the $Z$ resonance they find: $M_Z = 91.179\pm0.007$GeV/c$^2$.~\cite{pdg}

\subsection{Measurements of a Yukawa Coupling}

Next on the list of Standard Model parameters to measure is a fermion mass, or
if you prefer, a Yukawa coupling. The masses of all the fermions have been
measured (or in the case of the neutrinos, upper limits  have been set on their masses with
a possible lower limit coming from LANL observation of
$\nu$ oscillations~\cite{nuosc}).
The masses of charged leptons are determined quite precisely but
the masses of quarks are less well
known due to the complications of the strong interactions.
The masses are listed in Table~\ref{tab:mass}~\cite{pdg}.

\begin{table}[t]
\caption{Masses of the known fermions in MeV/c$^2$.
\label{tab:mass}}

\vspace{0.4cm}
\begin{center}
\begin{tabular}{|c|c|}
\hline
 & \\
$m_e$ = 0.51099906(15) & $m_{\nu_e}$  $<7 \times 10^{-6}$\\ 
 & \\\hline
 & \\
$m_\mu$ = 105.658389(34) & $m_{\nu_\mu}$   $<0.27$\\
 & \\ \hline
 & \\
$m_\tau$ = $1771.1 ^{+0.4}_{-0.5}$ & $m_{\nu_\tau}$  $< 31$\\
 & \\ \hline
 & \\
$ 2 < m_u < 8$  & $ 5 < m_d < 15$\\ 
 & \\ \hline
 & \\
$ 10^3 < m_c < 1.6 \times 10^3 $  & $10^2 < m_s < 3 \times 10^2$\\
 & \\ \hline
 & \\
$ m_t = (1.76\pm 0.16)\times 10^5$  &  $4.1 \times 10^3 < m_b < 4.5 \times 10^3$\\
 & \\
\hline
\end{tabular}
\end{center}
\end{table}

I will discuss the recent measurement of the top quark mass as an example of how
to measure a Yukawa coupling. Note that this is quite unique among the
fermion mass
measurements: the top quark is the highest rest mass particle ever observed!

To measure the top quark mass, you must first discover the top quark!
Actually, this is not true. The constraints on the top mass from the electroweak
radiative
corrections measured at the $Z$ pole were quite impressive. 
Before the top quark was discovered, Langacker and Erler~\cite{lang}
found $m_t =169 \pm 24$ GeV/c$^2$; however, lots of assumptions go into that
``measurement.'' The basic assumption is that there is nothing else new in the Standard Model
beyond what we already know.
However, I  want to talk about the measurement of the top
quark mass from the reconstruction of its decay
products.

\begin{figure}
%\vskip 4.5in
\centerline{\psfig{figure=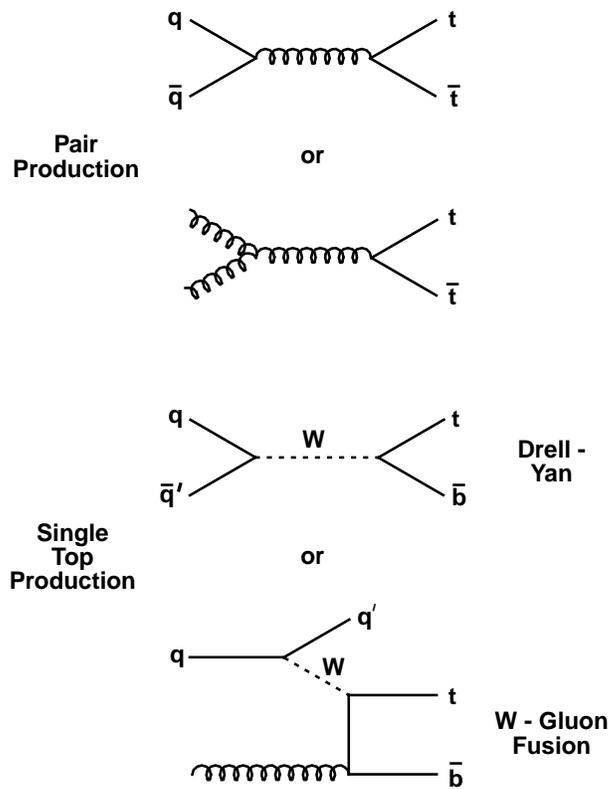,height=4.5in}}
\caption{The dominant top quark  production mechanisms at FNAL:  
$t\bar t$ pair production,  single $t$ production via Drell-Yan,
and single $t$ production in $W$-gluon fusion.
\label{fig:top}}
\end{figure}

With the mass of the top quark so high, the only machine capable of producing $t$ is the
Tevatron.  
Top quarks are produced in $p\bar p$ collisions by three main
mechanisms illustrated in Figure~\ref{fig:top}:
\begin{itemize}
\item $t\bar t$ pair production
\item single $t$ production via Drell-Yan
\item single $t$ production in $W$-gluon fusion.
\end{itemize}
At FNAL, for a high top quark mass, production is dominated by $t \bar t$ pair production.

\begin{figure}
%\vskip 3.00in
\centerline{\psfig{figure=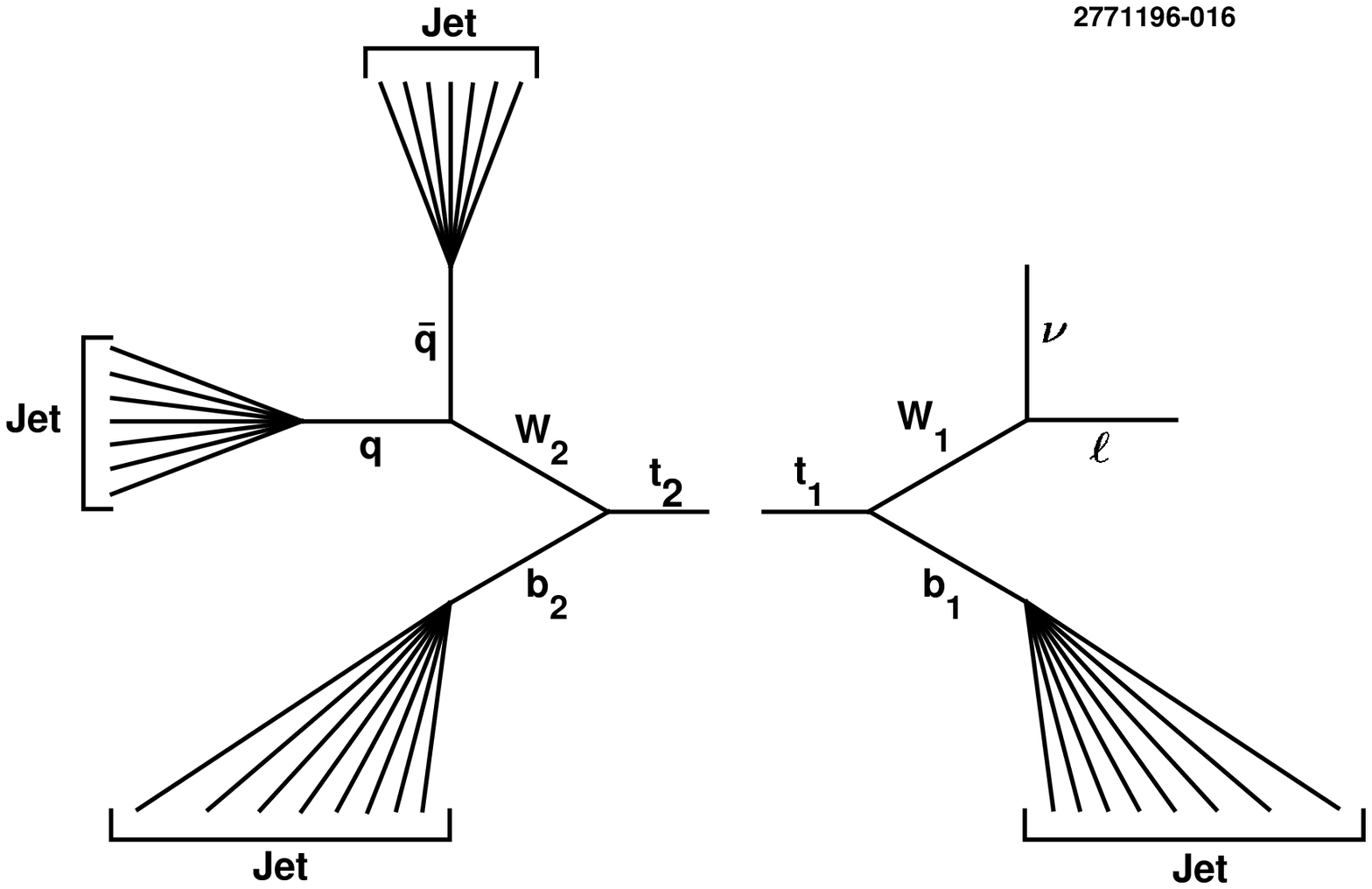,height=3.00in}}
\caption{A schematic view of the lepton plus jets decay mode of the two top
quarks produced at FNAL energies.  This is the channel used in determining
the top quark mass at CDF.
\label{fig:tchan}}
\end{figure}

The dominant decay of top is to $Wb$ followed by $W\to\ell\nu, u\bar d,$ or 
$c\bar s$.
There will be two $b$ quarks and 2 $W$'s in each event.  There are two classes
of $t \bar t$ events that can be reconstructed:  the first class is where both
$W$'s decay leptonically to electrons or muons (5\% of all decays), and the second class is where
one $W$ decays leptonically to an electron or a muon and the other $W$ decays
to two jets (30\% of all decays).  
For the initial discovery of top, both channels were used.  However, for the mass
measurement, only  a subset of the events was used.  The reason is that in the
case where both $W$'s decay leptonically, the system is underconstrained for
the mass measurement since there are two missing neutrinos in the event.
I will concentrate on describing the selection of $t\bar t$ events in the lepton
and jets channel, which is illustrated in Figure~\ref{fig:tchan}, since 
these are the events
used for the mass measurement~\cite{franklin}. 
 
A standard way to measure a particle mass is to measure the
momenta of all of the decay products and then
reconstruct the invariant mass of the parent particle.
For the top search, the implementation of this procedure is not
obvious. For events in which
\begin{equation} t_1\to W_1 b_1, \qquad W_1\to\ell\nu \qquad
t_2\to W_2 b_2, \qquad W_2\to q\bar q
\end{equation}
the momentum and energy of the lepton can be measured in a
straightforward fashion.  The measurement of the momenta and energies of the
quarks is hard.
This would all be much simpler if we could detect quarks, but we
cannot. Quarks hadronize, making clusters of particles in the detector called
jets.  The jet energy resolution is poor,  there can
be gluon radiation giving extra jets in the event,
 and the combinatorics are not favorable as
there
are
24 ways of assigning the four jets detected to the four final state quarks.  
Fortunately one can require 
one of the jets to be a $b$ jet in order
to reduce combinatorics.
The field of jet spectroscopy is in its infancy. As we go to higher
energy machines, we will have to get better at it!

With the jet  and lepton 4-vectors in hand there are
three constraints to calculate the remaining unknown neutrino 4-vector:
\begin{itemize}
\item $M\left(q\bar q\right) = M_W$
\item $M\left(\nu\ell\right) = M_W$ \item $m\left(t_1\right) = 
m\left(t_2\right)$
\end{itemize}

Events for the $t$ mass measurement in CDF are selected by requiring one hard 
lepton and four or more jets.  
CDF   requires that
one jet in the event have a $b$-tag: either there
is a separated vertex consistent with the finite $b$ lifetime,
 or a soft lepton consistent with coming
from a $B$ meson decay.
After the $b$-tag, CDF has a sample of 19 events,where approximately 6 are
estimated to come from background.

\begin{figure}
%\vskip 4.0in
\centerline{\psfig{figure=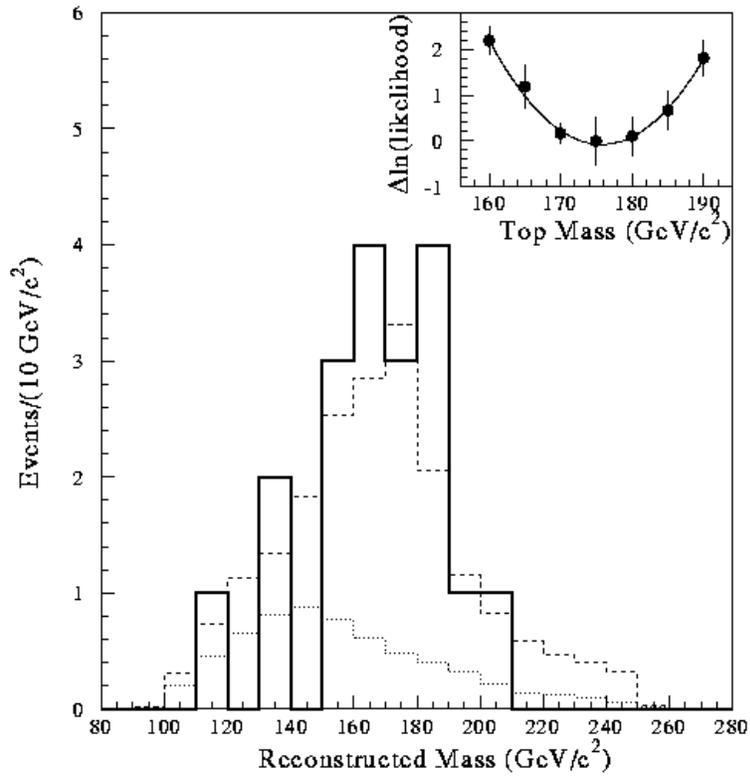,height=4.0in}}
\caption{The reconstructed mass distribution of the top quark
from CDF.  The solid line is the data.  The dotted line is the background
shape, and the sum of the background and $t \bar{t}$ Monte Carlo is
the dashed line.
\label{fig:cdftop}}
\end{figure}

The top quark mass is calculated
using the information from the
constrained fit.  If the jets are correctly assigned, the resolution on the
mass is  $\approx 12$GeV/c$^2$. Because of gluon radiation and
incorrect assignments, the resolution is in practice $\approx 24$GeV/c$^2$.
The CDF distribution of invariant mass formed from decay products is
shown in Figure~\ref{fig:cdftop} and gives $m_t =
176\pm8\pm10$, where the systematic error is
determined by the uncertainty in the response of the calorimeter to hadrons
giving an uncertainty in the jet energy scale, and the uncertainty in the
underlying QCD
processes that form the jets~\cite{cdftop}.

The top mass measurement is among the
first examples of using jet spectroscopy to determine the mass of a
particle.  We can expect to see this technique improved and exploited at future machines.

\subsection{Measuring a Mixing Angle}

The final experiment I want to discuss is how to measure a quark mixing angle. Recall
that the
quark eigenstates of the strong interactions, which
are the states of definite  flavor, are not eigenstates of the weak interactions.
In quantum mechanical terms, flavor is a symmetry of the strong interactions,
 so the strong interaction is diagonal on the quark flavor
basis ($u, c, t, d, s, b$).  However, the weak
interaction is diagonal on a different basis ($u, c, t, d',
s', b'$) and there is some unknown and undetermined rotation matrix that
relates the two bases.
It is up to experiment to determine the elements of the $3\times 3$ 
rotation matrix.
The effect of the flavor mixing is that the strength of the weak
interaction between two quark states of
definite flavor will be modified by a coefficient from this
rotation matrix to account for the flavor mixing.
For example, the strength (and hence the rate) of $b$ to $c$ decay will be modified by an
unknown factor we will call $V_{cb}$. By measuring the decay rate, we can extract
$V_{cb}$.

\begin{figure}
%\vskip 3.325in
\psfig{figure=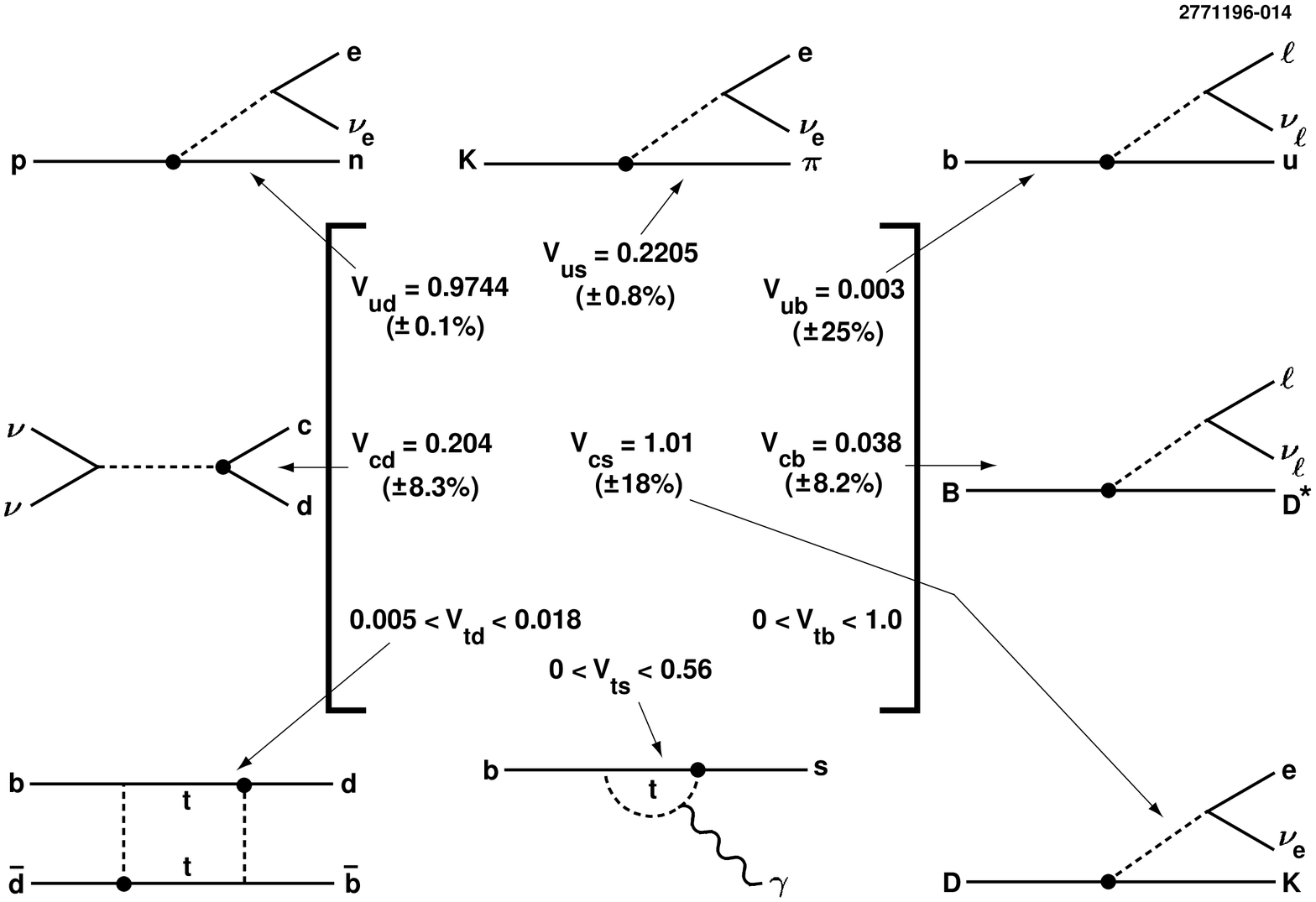,height=3.325in}
\caption{The measured values of the CKM matrix elements and a schematic
indication of the processes used to measure them.
\label{fig:ckm}}
\end{figure}

Figure~\ref{fig:ckm} gives a summary of the
CKM matrix and how the elements are measured~\cite{pdg}. The
values quoted are what are actually measured.  The requirement that
the matrix be unitary provides a powerful constraint on the poorly measured
elements.

As an example, I will
discuss a  measurement of $V_{cb}$ by the
CLEO detector using $B$ mesons produced by CESR.  Let
me remind you that CESR is an
electron-positron storage ring with
a center of mass energy $E_{CM} =
10.58$GeV. At that energy, one is just above threshold to produce a pair
of $B$ mesons and nothing else.

The strategy for the
measurement is to use the decay $B\to D^*\ell\nu$.
In Section 3.4, I discussed briefly how to select the event
sample, evaluate backgrounds and efficiencies, and measure the rate for the decay.
Recall that experimentally,  we measure a branching  ratio which is related to 
the decay rate
by the relation:
\begin{equation}
\Gamma\left(B\to D^*\ell\nu\right) = \frac{{\rm
BR}\left(B\to D^*\ell\nu\right)}{\tau_B}
\end{equation}
We use Fermi's golden rule to relate the
measured decay rate to the underlying physics:
\begin{equation}
\Gamma_{B\to D^*\ell\nu} = \frac{2\pi}{\hbar}|< D^*\ell\nu | V_{cb} {\cal H} | B > |^2
\rho\left(E\right) = \kappa V_{cb}^2
\end{equation}
We need to evaluate the matrix element (overlap integral)
in order to get the constant of proportionality between $\Gamma$ and $V_{cb}$ and here
we 
run into trouble because we cannot solve the
Schrodinger equation for the $B$ meson.  We do not know what
wave functions  to put in for the
mesons in order to calculate the matrix element.
For years, theorists have made educated guesses for meson wave functions and generated
calculations for $\kappa$. For educated guess read systematic error on $V_{cb}$!
Now, however, there is a better way.  This
is an example of an area where the developments in theory and experiment go hand
in hand.

A new theoretical approach to calculating the
form factors or overlap integrals  for exclusive semileptonic decays has
attracted a lot of attention in recent years. The basic idea is to notice that a $B$
meson or a charm meson (in both cases a light quark bound to a very heavy quark) looks a lot like the
hydrogen atom, which is a light $e^-$ bound to a heavy proton.  This
approach is called heavy quark effective theory (HQET)~\cite{hqet}. What can it buy in the
extraction of CKM matrix elements?
Recall that the $e^- $ wave function
in the hydrogen atom is independent of the mass and spin of the proton, up to
hyperfine corrections. We might guess that the
light quark part of the meson wave function should be independent
of the mass of the heavy quark up to hyperfine corrections of order $\Lambda_{QCD}/m_Q$,
where $m_Q$ is the mass of the heavy quark, and, like the proton in the H atom, 
the heavy quark should behave like a free particle.

The implication of the two previous statements is that the meson wave function
should factorize and
therefore so does the matrix element. When one calculates the amplitude for this decay, 
there
is a heavy quark part describing the decay of a free $b$ quark to a free $c$ quark,
which is calculable, and there is a light quark overlap integral describing the
probability for the light quark cloud in the initial state to turn into the light quark
cloud in the final state.  This overlap integral  depends on the
velocities of incoming and outgoing
mesons, and is not calculable from first principles.

However,
the light quark overlap integral is universal.  It
should be the same for all heavy pseudoscalar or vector meson to 
 heavy pseudoscalar or vector meson decays.  It is  called
the Isgur-Wise function:
$\xi(v\cdot v')$. All of the form factors for $B\to D, D^*\ell\nu$
decays can be written in terms of $\xi$.
It is a help that now there is only one unknown in the problem that needs to be
modeled, but this
is still an experimentally unsatisfactory situation since we know nothing about
$\xi$.

What makes HQET attractive is that at zero recoil, when the
initial and final state
mesons are at rest, $\xi\left(v\cdot v' = 1\right) = 1$, the form factors
describing the overlap of the initial and final light quark wave
functions are absolutely
normalized. This absolute normalization is the result of the fact that at zero
recoil, since the light quark wave function is
 independent of the mass of the heavy quark, the light quark
does not know a $c$ quark has replaced a $b$ quark. There is no velocity mismatch and
the overlap is perfect.
At this magic kinematic point, you can measure $V_{cb}$ independent of any
unknown form factor. Perhaps a simpler way of saying it is one can trade statistics in
data to measure the decay rate in a corner of phase space where the 
form factor is well known.

\begin{figure}
%\vskip 2.75in
\centerline{\psfig{figure=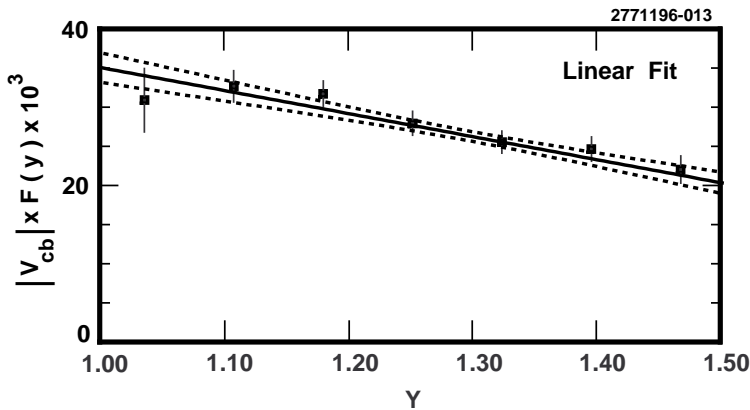,height=2.75in}}
\caption{The differential decay distribution for $B\to D^*\ell\nu$ decays
from CLEO.  The left intercept yields $V_{cb}$.  
\label{fig:d*lnu}}
\end{figure}

HQET is only an approximation. There are corrections to it. By a stroke of good
fortune, the point of zero recoil is protected from first order
corrections in $\Lambda_{QCD}/m_Q$ and we need only worry about second order effects.
We can extract $V_{cb}$ from $B\to D^*\ell\nu$ taking advantage of HQET by
plotting the differential branching ratio as a function of $y = v\cdot v'$.
HQET says that at zero recoil ($q_{\rm max}^2$ or $y=1$) the form factor is
1, up
to hyperfine corrections. Therefore, when properly normalized, the intercept of
the differential decay rate at $y=1$ 
yields $V_{cb}$.
To determine the intercept, the
data are extrapolated from $q^2 < q_{\rm max}$, using a
linear expansion of the Isgur Wise function as shown in 
Figure~\ref{fig:d*lnu}.  From this analysis, we find~\cite{cleovcb} $|V_{cb}| = 
0.0362 \pm 0.0019 \pm 0.0024$.

I would like to point out that to many people, form factor
 calculations may seem to simply be grubby
phenomenology, but in the extraction of the CKM matrix elements, 
we are in most cases
limited by our understanding of the hadronic matrix elements in our determination of these
fundamental parameters of the Standard Model!

\section{Testing the Standard Model}

\subsection{The Search for the Higgs}

There is one final element that we need in order to define the SM and it will 
also provide us with a profound test of the model.  In its simplest versions,
the SM predicts --  in fact requires -- the existence of a neutral scalar
particle: the Higgs boson.  The observation of the Higgs is the most important 
prediction of the SM that has not yet been verified by experiment.  It is a challenge
to search for the Higgs because while the coupling of the Higgs to the fermions and
gauge bosons is now completely determined by the experiments we have
discussed so far, we have no prediction of the Higgs mass from the theory.
It is hard to design experiments to search for the Higgs, since they must be
sensitive to all masses!

In practice, the search is not so difficult as it might at first seem.
The pre-LEP experiments were able to rule out different chunks of the mass
range.  The LEP I experiments~\cite{pdg} were able to eliminate a Higgs with $0 \leq M_H \leq
58.4$ GeV/c$^2$ and LEP II will extend that limit (assuming they don't find the
Higgs) to almost 100 GeV/c$^2$.  

There is another aspect of the Higgs search that needs discussing.  There are
very few in the high energy community who believe that the minimal
Standard Model is a fully satisfactory and 
complete description of nature.  The model contains many arbitrary parameters
and the mechanism for  giving mass to the fermions is totally ad hoc.  There
is a sense that there must be something more.  As a result, most high 
energy physicists view the search for the Higgs not as a final nail in the coffin
(once the Higgs is found and its mass measured, then we have a complete theory) but
rather the search for the Higgs is the most likely gateway to really new physics.

There is an excellent discussion in the text of Peskin and Schroeder~\cite{peskin}
that asks whether the $W$ and $Z$ might have acquired their mass by some
different, more complicated mechanism than spontaneous symmetry breaking, since
there is no experimental evidence for the Higgs.  Peskin and Schroeder argue
that there is compelling experimental evidence that the underlying theory
of the weak interactions is a spontaneously broken gauge theory.  There is
no other principle except for a spontaneously broken gauge theory that could
explain the experimental observation of universal, flavor-independent coupling
constants that describe the entire range of neutral current phenomena.  
However, they point
out that the mechanism of spontaneous breaking of $SU(2)_L \times U(1)$ could be
much more complicated than the simple model of a single scalar field.  The breaking might
be the result of the dynamics of a complicated new set of
particles and interactions:  a Higgs sector instead of a single Higgs particle.
This new sector would have to generate the masses of the $W$ and $Z$ bosons 
in the relation: $M_W = M_Z\cos\theta_W$, and must also generate the masses
of the quarks and leptons.  The experimental implications are that we
not only need to find the Higgs, we must be alert to the possibility of
an  entire spectrum of Higgs particles.  First however, we must find some
evidence that at least one neutral Higgs particle exists.

\subsubsection{Rates and Strategies}

\begin{figure}
%\vskip 2.0in
\centerline{\psfig{figure=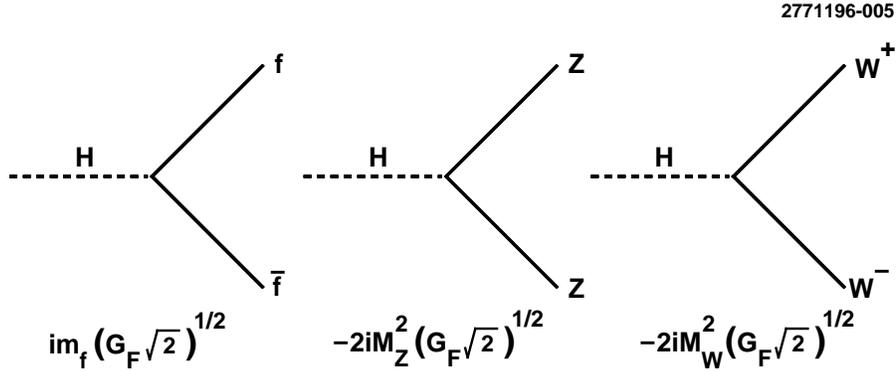,height=2.0in}}
\caption{Couplings of the Higgs to fermions and gauge bosons in the Standard
Model.
\label{fig:higgs}}
\end{figure}

It is easy to write down the SM couplings for the Higgs.~\cite{quigg}  The Higgs couples to
all fermions, $W$ and $Z$ bosons, and to itself, and all the couplings are
predicted as shown in Figure~\ref{fig:higgs}.
Because the Higgs coupling to fermions is proportional to the 
fermion mass divided by $M_W$, the production and detection of the Higgs is
difficult.  The Higgs just does not couple very strongly to stable matter.
If the Higgs mass is less than twice the $W$ or $Z$ mass, one will search for
the Higgs in final states involving the heaviest mass fermion available.  It is 
straightforward to calculate the decay rate:
\begin{eqnarray}
\Gamma(H\to f\bar{f}) & = & \frac{G_F M_H m_f^2}{4\pi\sqrt{2}}; M_H \gg m_f
\end{eqnarray}
where $m_f$ is the mass of the fermion in the final state.

The highly suppressed coupling of the Higgs to fermions makes it difficult to produce
the Higgs.  A first thought on how to search for a Higgs might be to
produce it in $e^+e^-$ collisions: $e^+e^- \to H \to f\bar{f}$ and look
for the resonance peak.  However,
the cross section for that process is tiny:
\begin{eqnarray}
\sigma(s = M_H^2) & \sim & 0.0005 {\rm nb}
\end{eqnarray}
for a 10 GeV Higgs and this is to be compared with the cross section for
producing quarks at the energy which is 3.65 nb.  The resonance bump would
not be experimentally observable.

A much more promising way to search for the Higgs is to take advantage of the
large coupling of the Higgs to the gauge bosons in order to produce the Higgs.
If the Higgs is heavy enough ($M_H \geq 2M_W$) one can also use gauge bosons
in the final state to search for the Higgs.

\subsubsection{Higgs Searches at LEP}

\begin{figure}
%\vskip 2.5in
\centerline{\psfig{figure=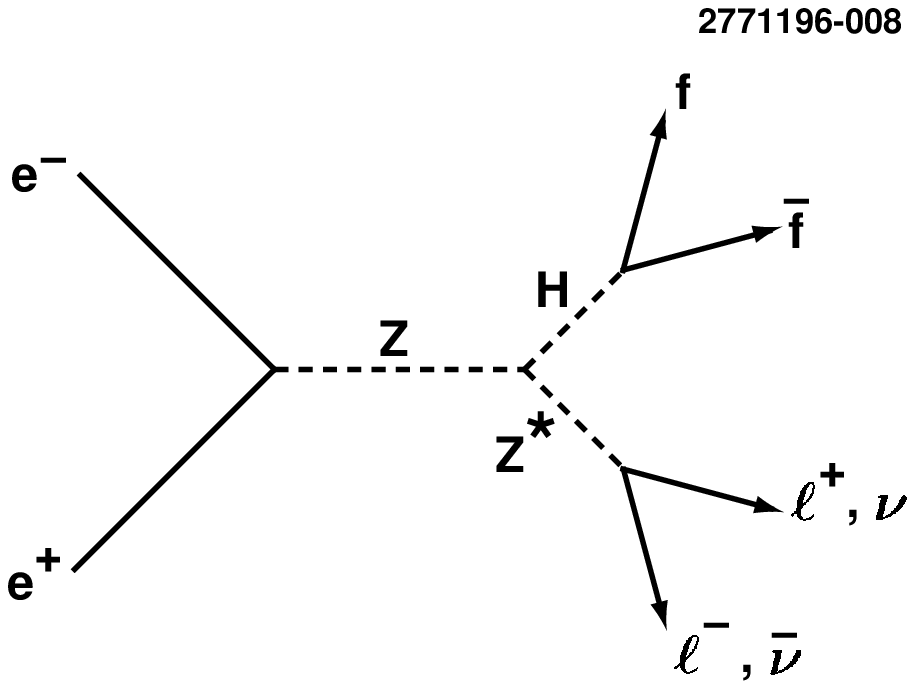,height=2.5in}}
\caption{Higgs production in $e^+e^-$ collisions at the $Z$ pole.
\label{fig:lephiggs}}
\end{figure}

There were experiments that gave somewhat model dependent limits on the neutral
Higgs mass before LEP, but the LEP searches are the most comprehensive.  The
way that the LEP experiments search for the Higgs is the bremsstrahlung
process illustrated in Figure~\ref{fig:lephiggs}.  The search is 
optimized as a function of the mass of the Higgs and since all the
LEP searches are only sensitive to
Higgs masses where $M_H \leq M_Z$, they look for decays to fermions in the final
state.  The searches all take advantage of the fairly unique topology of the
events illustrated in Figure~\ref{fig:lephiggs}:  a Higgs decaying to a 
fermion pair or hadrons recoiling against a $Z$ that decays to leptons or
neutrinos.  No events have been observed in any of the searches at the
$Z$ pole where the LEP I collider was operating at $E_{CM} = 92$ GeV leading
to a lower limit on the Higgs mass~\cite{pdg}:
\begin{eqnarray}
m_H & \geq& 58.4 {\rm GeV/c}^2
\end{eqnarray}
  
What are our prospects for extending the mass reach in our search for the
Higgs?  The LEP center of mass energy is currently being extended 
(LEP II) with
the addition of more RF cavities in the machine.  Currently LEP is operating
at 161 GeV and the ultimate goal is 192 GeV.  By looking for
$e^+e^-\to ZH$, a neutral Higgs can be discovered up to a mass of
$M_H \sim 0.97(E_{CM} - M_Z) \sim 95$ GeV/c$^2$.  If the neutral Higgs is
not found at LEP II, we will probably need to wait until LHC.

\subsubsection{High Mass Higgs Searches at LHC}

If we speculate that LEP II does not discover a neutral Higgs and puts a lower
limit on its mass of 95 GeV/c$^2$, what is next?  The next machine on the high
energy frontier (as it is often poetically called) is the Large Hadron
Collider: LHC.  The LHC will collide 7 TeV protons on 7 TeV protons 
in the LEP tunnel.  The design luminosity of the machine is $1 \times 10^{34}$
cm$^{-2}$s$^{-1}$, higher than any storage ring ever built, and searching
for the Higgs is one of the design goals of the experiment.  There will be
two detectors operating at LHC: CMS and ATLAS.  The detectors are being
optimized to be sensitive to the existence of a Higgs to the full kinematic
range of the machine (a mass reach of about 1TeV/c$^2$).  Let's speculate
a bit on how to search for the Higgs at LHC, if the Higgs has not been 
discovered before LHC turns on (around 2005).  

\begin{figure}
%\vskip 4.0in
\centerline{\psfig{figure=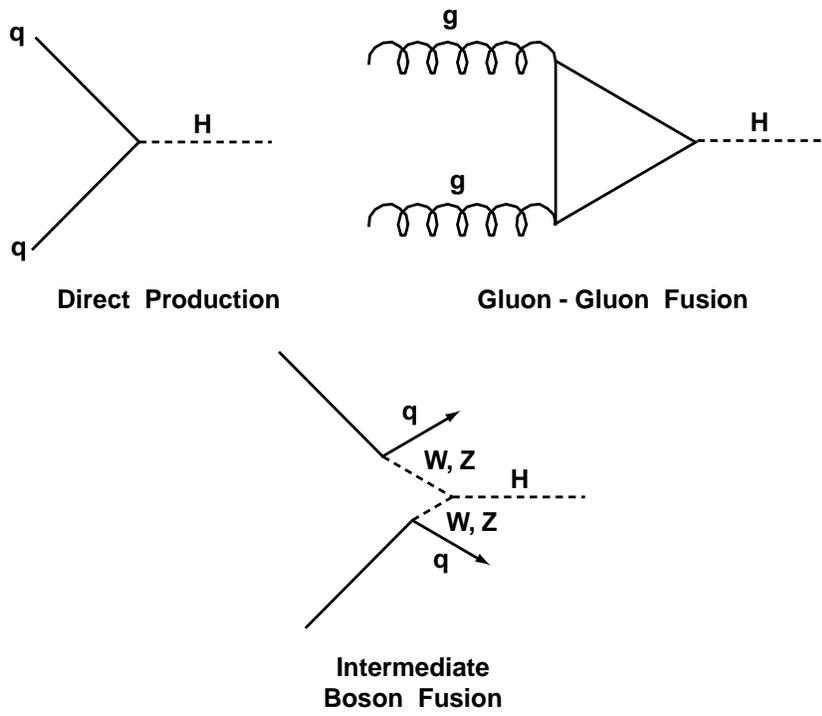,height=4.0in}}
\caption{Higgs production in $pp$ collisions at the LHC.
\label{fig:lhchiggs}}
\end{figure}

The production mechanisms for the Higgs at a proton-proton collider are
direct production, gluon-gluon fusion or intermediate boson fusion, as
illustrated in Figure~\ref{fig:lhchiggs}.

The search strategy for finding the Higgs at
LHC depends on the Higgs mass.  At high mass ($M_H \geq 130$GeV/c$^2$) the 
search is straightforward.  One looks for $pp \to H \to ZZ^* ~{\rm or}~ ZZ$
with the $Z$'s decaying to leptons.  The final state is clean and backgrounds
seem manageable.  The main issue is luminosity, particularly as one goes
to higher mass and the natural width of the Higgs becomes large.  One
needs 10 fb$^{-1}$ of data (one year at 1/10 design luminosity) to
discover a 500 GeV/c$^2$ Higgs in this channel, and 100 fb$^{-1}$ of data 
(one year at full design luminosity) to
discover an 800 GeV/c$^2$ Higgs~\cite{lhchiggs}.

The most challenging problem facing LHC experiments is to detect a Higgs in
the intermediate mass range: 95 GeV/c$^2 \leq M_H \leq 130$GeV/c$^2$.
The dominant decay mode of the Higgs in this mass range will be to two
$b$ quarks; a final state for which there will be an enormous QCD background.
The rare decay $H \to \gamma\gamma$ is the only hope in this region and even
that will be tough.  One needs good electromagnetic calorimetry and a lot of
data.

I want to stress that it is extremely important that we cover the entire
range of possible Higgs mass completely and convincingly.  Any hole of 10
or 20 GeV/c$^2$ could be fatal since that is where the Higgs might be
hiding!

I have discussed the search for Higgs at some length because the existence
of the Higgs is, as stated earlier, the most important prediction of the SM
that has not been verified.  It is also where, if there are to be answers to our
questioning of whether the minimal SM is really it, we hope to have the
first hints of answers.  Most extensions of the SM have a Higgs sector.  For
example, the minimal supersymmetric SM has two Higgs doublets, with 
one charged Higgs, two neutral scalars and one pseudoscalar, all waiting to be
discovered.  However, searching for the Higgs is not the only way to look
for new physics.

\subsection{Precision Tests of the Standard Model at the $Z$ Pole}

The most stringent tests of the SM are the elegant series of
analyses being done at LEP and SLC to make precision measurements of
SM observables at the $Z$ pole.  These experiments 
offer less hope to discover new physics than Higgs searches, 
since if discrepancies with the SM turn up, we may have to wait for the higher
energy machines to resolve the discrepancies and uncover the new physics
that causes them.  In a crude sense, precision experiments can tell us
if something is wrong with a theory, but they can't necessarily tell us
the cause.

In Section 4.2, I discussed the measurement of the $Z$ mass.
However, the LEP experiments can do a great deal more with the
approximately 6 million $Z$ decays they have accumulated.  They can do 
precision studies of how the $Z$ decays, and make detailed measurements
of the coupling of the $Z$ to the matter fields of quarks and leptons.  

For the lepton and $b$ and $c$ quark flavor
final states, one can count the number of decays to
an exclusive lepton or quark pair which, 
for each species, is proportional to the 
squares of the axial and vector coupling of the fermion or
quark to the $Z$.  One
can  form forward-backward asymmetries, where one keeps track of the
direction of the fermion or quark with respect to the incident electron
direction, and  at SLC, where the incoming electron
beam can be polarized, one can also form right-left asymmetries from the
cross section measured with right or left handed incoming electrons.
The asymmetries are proportional
to the product of the vector and axial coupling of the fermion or
quark to the $Z$.  
At tree level, all the asymmetries and partial widths
are completely determined from $G_F, \alpha_{EM}, \alpha_s$ and
$M_Z$, all of which are now well measured.  The one loop 
electroweak radiative corrections depend on $m_t$ and $M_H$, as well
as any possible new physics.  What can we do with the wealth of precision
measurements available?

\begin{figure}
%\vskip 5.5in
\centerline{\psfig{figure=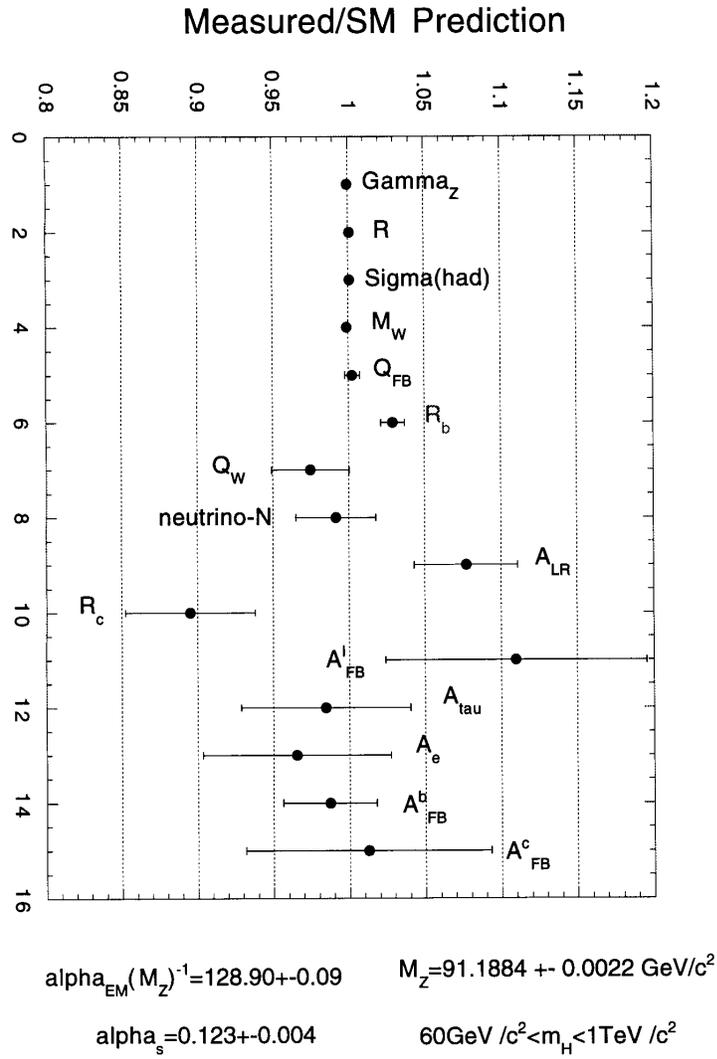,height=5.5in}}
\caption{Comparison of experimental measurements with Standard Model
predictions for the precision electroweak observables.  The input values
for the Standard Model parameters are shown.
\label{fig:sm}}
\end{figure}

The most straightforward thing to do with all the LEP/SLC precision 
measurements is to input the
best experimental values for $G_F, \alpha_{EM}, \alpha_s, M_Z$,
$m_t$,  and some reasonable range of values for $M_H$ (such as
60 GeV/c$^2 \leq M_H \leq 1 $TeV/c$^2$) and then see if the predictions of
the SM can be verified on a case by case basis with the precision
measurements of asymmetries and partial widths.  The results of this
approach are shown in Figure~\ref{fig:sm}, where I have plotted the ratio
of the SM value for the experimental quantities to the SM prediction~\cite{pdg}.
The points include LEP data, the $W$ mass measured at  FNAL,
and the deep inelastic neutrino scattering results.  We see in
Figure~\ref{fig:sm} that the SM is doing amazingly well!

There is only one result that was causing concern at the time I gave
these lectures and that was the measurement of $R_b$, the partial width
for the $Z$ to decay to $b$ quarks, which was almost four sigma above its
predicted value.

We can now think of doing the exercise with the Higgs that we did with
the top quark, and use these precision results to ``measure" the Higgs
mass through the effect that $M_H$ exerts on the radiative corrections to these
observables.  In order to do this, we
will have to assume that there is no new physics beyond
the SM.  The exercise is more difficult than in Section 4.2 where we used
the precision measurements to determine the top quark mass.  In
that case, $\delta r$ depended quadratically on $m_t$, 
but it depends only logarithmically on $M_H$.  Using the data in
Figure~\ref{fig:sm} one finds $M_H \leq 320(430)$ GeV/c$^2$ at 90(95)\%
confidence level but the constraint of a low Higgs mass is driven largely
by the measurements of $R_b$ and $A_{LR}$ which show, in this
data, slight discrepancies with the SM predictions~\cite{pdg}. I personally don't take
the limits on the Higgs mass from this kind of global fit very seriously.
I would need to be convinced that the discrepancies in
$R_b$ and $A_{LR}$ are real before buying stock in a low mass Higgs.

A preferable way to test the SM with the precision electroweak measurements
is to try to see if one can limit possible deviations from the SM in some model 
independent way.  There are several schemes that are popular for doing this.
Their authors usually consider the general effects on the
neutral current  and the $Z$ and $W$ pole observables of new heavy physics which contributes
to the $Z$ and $W$ self energies but doesn't have direct coupling to 
ordinary fermions.  

\begin{figure}
%\vskip 3.0in
\centerline{\psfig{figure=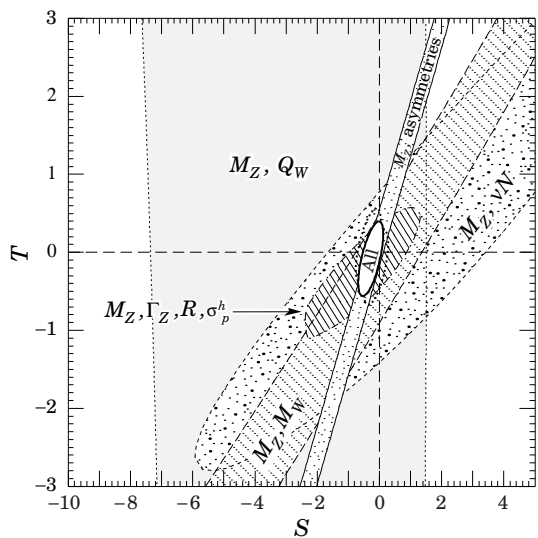,height=2.75in}}
\caption{Constraints on extensions of the Standard Model from the precision
electroweak data. 
\label{fig:stu}}
\end{figure}

One such scheme that is popular with experimentalists is the $STU$ scheme
of Peskin and Takeuchi~\cite{stu}, where the effects of new physics are parameterized in terms
of changes of the boson self energies:
\begin{itemize}
\item T is proportional to the difference between the $W$ and $Z$ self energies
at $Q^2 = 0$.
\item S is the difference between the $Z$ self energy at $Q^2 = M_Z^2$ and 
$Q^2 = 0$.
\item S+U is the difference between the $W$ self energy at $Q^2 = M_Z^2$ and 
$Q^2 = 0$.
\end{itemize}
The interested reader can find this scheme described in more detail in
the references~\cite{stu}.  Most of the electroweak observables depend on $S$ and
$T$ only, and in the SM, $S=T=U=0$.  Figure~\ref{fig:stu}~\cite{pdg} shows all the
data on the $S-T$ plane and we see that all of the measurements are consistent
with no new physics.  The contours drawn assume
$M_H = 300$ GeV/c$^2$, with the exception of the
two contours for all data which are displaced
slightly upward (downward) corresponding to 
$M_H = 1000 (60)$ GeV/c$^2$.  We see that while the lightest mass allowed
for the Higgs is favored, the data are not yet very sensitive to $M_H$ and the
full range on the Higgs mass from $60$ GeV/$c^2$ to 1 TeV/$c^2$ is still
allowed.

\subsection{Other Tests of the Standard Model and Searches for New Physics}
I have certainly not exhausted all of the tests of the SM that are going on.  The two
I have discussed so far are the most focused, but there are a slew of other experiments
that are looking for processes forbidden in the SM or very highly suppressed, and whose 
appearance at an unexpectedly high level might be a hint of something new!

Examples of such experiments are:
\begin{itemize}
\item the search for flavor changing neutral currents
\item searches for lepton number violation 
\item tests of CPT invariance
\item limits on proton decay
\end{itemize}
and many others.  These tests of the SM have proved extremely useful in constraining new
physics proposed by theorists.  Recently, however, they have not produced any surprises.
Nevertheless, it is important to keep looking.

\subsection{The Status of the Standard Model}

As we have seen, the SM has proved very robust, and it has proved frustratingly successful
in predicting experimental results.  When I gave these lectures at TASI in June of
1996, I was able to point to a few experimental results that were slightly out of line
with theoretical predictions.  Keeping in mind the warning that, to date, every experiment
has either confirmed the SM predictions or the experiment has been wrong, I discussed
the ``rogue" results.  By the end of the summer, however, all of the discrepancies we
were seeing in June had gone away.  

\subsubsection{R$_b$}
As can be seen in Figure~\ref{fig:sm}, the measured value of $R_b$, the fraction of time the
$Z$ decays to a pair of $b$ quarks, was almost four standard deviations above its predicted
value.  The value used in the figure comes from taking the results published by the
four LEP experiments and averaging them~\cite{pdg}.  However, at the 1996 summer conferences, two
new measurements of $R_b$ were announced by the ALEPH and SLC 
collaborations~\cite{rb}
which were both more in line with the SM expectation, and the problem seems to be going 
away.

\subsubsection{Quark Substructure}
Last year the CDF collaboration published a result that could be interpreted as evidence
for quark substructure~\cite{cdf}.  They saw an excess of jets at high transverse energy in their
jet differential cross section distribution when compared with the expectations of
next to leading order QCD.  However, the other detector operating at the Tevatron, D0,
did not see a similar excess, and the theoretical errors associated with the parton
distribution functions are  probably larger than originally thought, and so this problem
also seems to be going away~\cite{d0}.

\subsubsection{ALEPH Four Jet Excess}
Another interesting result reported
in 1996  that was hard to explain within the context of the SM was
an excess of four jet events reported by the ALEPH collaboration from a supersymmetry
search looking for $e^+e^- \to hA$ in their high energy (130-136 GeV center of mass)
data~\cite{alephhiggs}.  They found an excess of events (16 events on a background of 8.6) at a
dijet mass of 105 GeV/c$^2$.  However, none of the other LEP experiments could reproduce
the result and the peak in the ALEPH data appears to have been a 
statistical fluctuation~\cite{otherhiggs}.

\subsubsection{Neutrino Oscillations}
An experiment that is worth keeping an eye on is the recent report of evidence for 
neutrino
oscillations from Los Alamos~\cite{nuosc}.  The SM has no problem with
a neutrino mass (in fact it seems unnatural for the neutrino to be exactly massless).  If
neutrinos do have a mass then it is expected that the lepton numbers won't be separately 
conserved and there will be lepton
mixing analogous to the quark flavor mixing already observed. 

The Los Alamos experiment takes a beam of $\bar{\nu}_\mu$ produced by
positive muons that are made by protons on a 
water target making pions and kaons which are then charge selected.  The positive
muons decay via $\mu^+ \to e^+ \nu_e \bar{\nu}_\mu$.  The experiment searches 
for $\bar{\nu}_e$ downstream of the  $\bar{\nu}_\mu$ beam 
(the result of a $\bar{\nu}_\mu$ to $\bar{\nu}_e$ oscillation)
via the reaction  $\bar{\nu}_e
+ p \to e^+ + n$.  They see 22  $\bar{\nu}_e$ events with an expected background of $4.6\pm0.6$
events.  If confirmed, this fills out the SM in a rather attractive way, but any future
explanation for quark mixing will have to explain lepton mixing as well.

\subsection{The Future}
I want to end with a peek into the future.  What will our reach be in terms of 
probing the SM in the next decades?  Where will the action be?  What will we hope to
learn?

\subsubsection{The Near Term Future}
In the very near term we will be hearing from FNAL and LEP II.  LEP II is pushing to the
highest energy $e^+e^-$ center of mass energy ever achieved.  They are currently running
at 161 GeV and have an eventual goal of about 195 GeV.  They will put the best limits on the
SM Higgs mass until well into the next decade.

FNAL will be upgrading its luminosity by a factor of 10 by the end of the decade.  This will
allow them to have a much larger top sample and they will be the only machine looking
at top until well into the next decade.  FNAL will also have the highest energy reach
of any machine, and the upgrade deepens that reach as the increased luminosity
helps populate the high energy tails of the parton energy distributions.  They will
have the best chance of discovering supersymmetric particles such as squarks and
gluinos.

The other area of intense experimental activity in the next decade is
at the B-factories.  These are $e^+e^-$ machines operating at a center of mass energy
of 10.58 GeV that will explore CP violation in $B$ mesons and hopefully nail down the
parameters of quark mixing.  

\subsubsection{On the Horizon:  The LHC}
The next big machine will be the Large Hadron Collider: LHC.  This is a proton-proton
collider that will run at CERN starting around 2005.  The machine and detectors
are being designed to have sensitivity to the largest possible Higgs mass range, since probing the
origins of mass at the electroweak scale is a major focus of interest.  They will also be
able to probe supersymmetry since gluinos and squarks will
be copiously produced (if they exist).  With a mass reach of about 1.5 TeV in the
supersymmetric particle searches, it is very likely that LHC will be able to confirm or
exclude the existence of supersymmetry.
The LHC has been approved and is under construction.  There are two big detector 
collaborations formed (CMS and ATLAS) and detector designs are well along.  

\subsubsection{Over the Horizon:  What is Next After LHC?} 

A serious
question confronting the experimental community is: what is next?  What is the next
machine after the LHC that we will want to build?  I don't know the answer to that
question.  I will present some of the ideas and proposals being discussed.  I want to alert
you to this debate because the outcome will profoundly affect {\it your} future!  This is
the beginning of the process that will decide what data will be coming in during your
career lifetime.  The LHC and this next machine will dominate the
first 20 or more years of the next century.  It is a crucial decision.

The most serious option for a next machine after the LHC is an $e^+e^-$ linear 
collider~\cite{nlc}.
This is a machine that would start with a center of mass energy of 500 GeV and might eventually be upgraded to 1.5 TeV.
If the minimal supersymmetric Standard Model is right,
this machine could be a gold mine 
 since most
of the Higgses and superpartners would be accessible even at 500 GeV center of mass 
energy.  While the discovery of SUSY might still take place at LHC, the $e^+e^-$ linear collider
offers a much cleaner experimental environment to fully study the SUSY particle spectrum.
If, however, the minimal supersymmetric model is wrong, then this machine needs to push to the
highest center of mass energies of over 1 TeV in order to extend our physics reach
past what already will be  learned from the LHC.

Another option that is being discussed as a future machine  is an even higher energy
proton-proton collider, with 100 TeV of energy in the center of mass.  This would be
a frontier machine that would push to the highest possible energy just to see what is
there~\cite{superpp}.

A final option being discussed is a $\mu^+\mu^-$ collider.  One wins with muons over
electrons since electron colliders are limited by radiation which is much suppressed
for muons because of their larger mass.  The talk is of a muon collider with 4 TeV
in the center of mass.~\cite{muoncollider}  

At this point, I don't know what direction the field will go and I don't know when a 
decision will be made about what machine will be built after the LHC.  My message that I
want to leave the reader with is that you should care which of these machines 
is built because the choice will determine the experimental results that will be available
during your career.

I would also like to remind the reader that it is important to keep an open
mind.  I suspect
that we know less about the way the world works than we think we do.  I don't know if the
critical data is already in our hands, whether it will come from a high energy machine,
or whether it will come from an unexpected direction such as proton decay, neutrino
oscillations, or high energy cosmic rays, but I do believe that we are  still in for some
surprises!

\section*{Acknowledgments}
It is a pleasure to thank Ken Bloom, Dave Crowcroft and Andy Foland for
their generous help and advice in preparing these lectures.  This work
was supported in part by the National Science Foundation.

\end{document}